\documentclass[a4paper,fleqn,usenatbib]{mnras}
\usepackage[T1]{fontenc}
\usepackage{ae,aecompl}
\usepackage{amsmath}
\usepackage{amsfonts}
\usepackage{amssymb}
\usepackage{graphicx}
\usepackage{color,xcolor}
\def\mnras{MNRAS}
\def\aap{A$\&$A}
\def\nat{Nature}
\def\apj{ApJ}
\def\apjl{ApJ Letters}
\def\apjs{ApJS}

\def\aj{AJ}
\def\icarus{Icarus}

\newcommand{\me}{\, {\rm M}_{\oplus}}
\newcommand{\msun}{\, {\rm M}_{\odot}}
\newcommand{\msunyr}{\, {\rm M}_{\odot}\,{\rm yr^{-1}}}
\newcommand{\rsun}{\, {\rm R}_{\odot}}
\newcommand{\au}{\, {\rm au}}

\newcommand{\Stokes}{{\, {\rm St}}}
\defcitealias{Coleman21}{Paper I}

\title[Wide Binaries vs Photoevaporation]{From dust to planets - II. Effects of wide binary companions and external photoevaporation on planetesimal and embryo formation}
\author[G. A. L. Coleman]{Gavin A. L. Coleman\thanks{Email: gavin.coleman@qmul.ac.uk}\\
Astronomy Unit, Department of Physics and Astronomy, Queen Mary University of London, Mile End Road, London, E1 4NS, UK}
\date{Accepted 2026 March 25. Received 2026 March 25; in original form 2025 October 28}
\pubyear{2026}
\begin{document}
\label{firstpage}
\pagerange{\pageref{firstpage}--\pageref{lastpage}}
\maketitle
\begin{abstract}
More than half of Solar-type stars are found in binary systems. The numbers of exoplanets within binary systems in s-type orbits now numbers over 700. However, whilst the numbers have increased, there still does not exist a global model of planet formation for wide binary systems, where there does for single stars and circumbinary systems. As a precursor to such a model, that includes the necessary physical planet formation processes, it is important to understand how an outer binary companion affects the evolution of circumstellar discs, and the formation of planetesimals and planetary embryos. The main mechanism for which these processes are affected, is through truncation of the protoplanetary disc outer edges. In this paper, we determine these effects, whilst also comparing them to the effects of external photoevaporation that competes to truncate protoplanetary discs. We find that disc truncation by both a binary companion and external photoevaporation significantly reduces the efficiency to which planetary embryos are able to accrete pebbles and grow into terrestrial mass planets. This is due to the pebble supply being cut off as the pebble production front reaches the disc outer edge before planets are able to significantly increase in mass. This hindrance to planet formation occurs when the truncation radius due to the binary companion is below $\sim 30\au$, corresponding to binary separations of $\sim90\au$ for equal mass, circular binary stars. For separations greater than 300$\au$, planet formation operates similar to that around single stars. Our results highlight the detrimental effects of a binary companion for intermediate binaries, that can provide possible explanations for the dearth of multiple planets within binary systems of separations $<100\au$.
\end{abstract}
\begin{keywords}
planets and satellites: formation -- protoplanetary discs -- binaries: general.
\end{keywords}

\section{Introduction}
\label{sec:intro}

To date, over 7000 planets have been discovered through a variety of methods, ranging from transit detection to direct imaging to radial velocity signatures (RV). Most of the planets that have been discovered are in single star systems \citep[see][for a review]{Winn15}, however a large number have been observed to be in systems with multiple stars. These include both those in circumbinary configurations where the planets orbit both stars (e.g. Kepler-16b \citep{Doyle11}, BEBOP-1c \citep{Standing23}), as well as those in circumstellar s-type orbits where the planets orbit a single star whilst one or more companions orbit further away (e.g. $\gamma$ Cephei Ab \citep{Campbell88,Hatzes03}, Proxima Centauri b \citep{Anglada2016}). Recent analysis has found that 759 of the known exoplanet systems contain multiple stars, 31 of which being circumbinary, and the remaining 728 being of the s-type configuration as either wide binaries, or triple/quadruple star systems \citep{Thebault25}. 

Whilst only $\sim10\%$ of known planets are found in multiple star systems, it is known that stellar multiplicity rates are actually larger. For Solar-type stars, 60$\%$ are found to be in multiple star systems \citep{Raghavan10}, whilst for late M-type stars only 10$\%$ are in multiples \citep{Offner23}. The differences between the multiplicity rates and the the percentage of multiple star systems known to host planets can be due to a number of factors. These include observational biases since RV surveys typically exclude systems where the companion is within 1--2'' of the primary, since the resultant RV signals are polluted and thus making it difficult to disentangle the individual spectra \citep{Eggenberger07,Hirsch21}. Recent work has however shown that radial velocities from double-lined binaries can be disentangled allowing planet detections to take place in circumbinary systems \citep{Sairam24}, though this has yet to occur for wider binary systems. As well as radial velocity observational biases, transit detections are reduced compared to single stars, mainly due to the additional flux of the companion stars acting to decrease the transit depth of planets that forces them to sink below the required signal-to-noise threshold for detection \citep{Wang15b,Ziegler20}. In addition to observation biases, there could also be intrinsic differences in the planet populations that exist in wide binary systems, compared to that of single stars.

As well as comparing the overall exoplanet occurrence rates within single and multiple star systems, other works have also attempted to compare the occurrence rates in single systems to that in multiple systems with different binary separations. One such study by \citet{Hirsch21} found that the giant planet occurrence rate for binaries of separations $>100\au$ seemed to be compatible with that for singles, whilst for closer binaries, the occurrence rates were significantly reduced. Another statistical study of s-type planets found through RVs also supports the idea that the planet populations around wide binaries of separations 100--300$\au$ are mostly similar to those found in single star systems \citep{Su21}. Additionally, \citet{Su21} also find that for smaller binary separations, there is a dearth of multiple planet systems consistent with that found in previous studies \citep{Roell12}, and an overabundance of close in giant planets. They suggest that enhanced planet migration, collisions and/or ejections in close binaries could explain most of the statistical patterns they observe in binary systems.

Previous models of planet formation for wide binary systems have mainly focussed on specific systems, such as Alpha Centauri \citep{Barbieri02,Quintana02,Thebault08,Thebault09}, and found that outer companions have numerous effects on planet formation processes. Firstly, perturbations from the outer companion act to tidally truncate protoplanetary discs to typically between 1/4 to 1/2 of the binary separation, depending on the binary parameters \citep{Papaloizou77,Artymowicz94,Savonije94}. This truncation decreases the amount of material in protoplanetary discs \citep{Jang-Condell08,Harris12} as well as reducing their lifetime \citep{Kraus12,Rosotti18}. In addition to the effects of the outer companion on the protoplanetary discs, gravitational perturbations from the companion also act to stir up planetesimal and planet eccentricities as they form and grow in their natal discs. This stirring increases the collision velocities between planetesimals, leading to more eroding collisions, and reducing their levels of mutual accretion \citep{Marzari00,Paardekooper08,Thebault04,Thebault06}. This was shown to particularly be the case for close binaries such as $\gamma$ Cephei or Alpha Centauri \citep{Paardekooper08,Thebault04,Thebault09}. All of these effects act to reduce the efficiency of planetary growth.

Whilst there have been a number of works that have explored the effects of a wide binary companion on planetesimal accretion for specific systems, and for the evolution of the protoplanetary discs, no global models of planet formation for wide binary systems, including the relevant processes, have been developed. Such models have been formulated for single stars \citep{ColemanNelson14,Mordasini15,Emsenhuber21a} and for circumbinary systems \citep{Coleman23,Coleman24FFP}, and are regularly compared with observations \citep{Emsenhuber25} and for explaining the formation of specific planetary systems \citep{ColemanProxima17,Coleman19,Standing23}. As a precursor to developing a global model of planet formation for wide binary systems, here we explore the effects of an outer companion on the formation efficiency of planetesimals and planetary embryos, as well as the efficiency of pebble accretion. Additionally we contrast and combine these effects to those that arise from nearby massive stars irradiating the discs with high energy radiation driving external photoevaporative winds \citep{Odell94,Adams04,Haworth19,Haworth23}, determining where either mechanism dominates the evolution of protoplanetary discs, and where their effects on the formation and growth of planetary embryos is sufficient to substantially frustrate the outcomes of planet formation.

This paper is laid out as follows. We outline our physical model in Sect. \ref{sec:base_model}, whilst we provide examples for the evolution of protoplanetary discs in different environments in Sect. \ref{sec:disc_evol}. We then explore the role of the pebble cutoff time in Sect. \ref{sec:peb_cutoff}. In Sect. \ref{sec:comp} we compare the effects of external photoevaporation to those arising due to an outer binary companion. Finally, in Sect. \ref{sec:conc} we draw our conclusions.

\section{Physical Model}
\label{sec:base_model}

Protoplanetary discs lose mass through accretion on to the central star and by photoevaporative winds launched from their surface layers. Typically, the outer edges of protoplanetary discs are regulated through the balance of viscous spreading and the loss of disc material through external photoevaporation \citep[e.g.][]{Coleman22,Coleman24MHD}. However, most studies have only been carried out for single stars, rather than for wide binary systems, where tidal torques from an outer companion act as a truncation mechanism for protoplanetary discs \citep{Papaloizou77,Artymowicz94}. For disc material interior to the truncation location for binary systems, it is found to orbit similarly to that of material around a single star \citep{Rosotti18,Zagaria21}. With this in mind, this allows us to treat the evolution of protoplanetary discs around each star within a binary system independently. Below we will describe how we model the protoplanetary discs within wide binary systems, taking into account accretion through the disc, photoevaporation, and the truncation of the companion stars.

\subsection{Disc Model}
For the underlying gas disc model, we utilise the model used in previous works \citep{Coleman22,Coleman24MHD,Coleman25}, that uses a 1D viscous disc model where the evolution of the gas surface density $\Sigma$ is solved through the standard diffusion equation

\begin{equation}
    \dot{\Sigma}(r)=\dfrac{1}{r}\dfrac{d}{dr}\left[3r^{1/2}\dfrac{d}{dr}\left(\nu\Sigma r^{1/2}\right)\right]-\dot{\Sigma}_{\rm PE}(r)
\end{equation}
where $\nu=\alpha H^2\Omega$ is the disc viscosity with viscous parameter $\alpha$ \citep{Shak}, $H$ the disc scale height, $\Omega$ the Keplerian frequency, and $\dot{\Sigma}_{\rm PE}(r)$ is the rate of change in surface density due to photoevaporative winds. To account for the role of the outer binary companion, as in \citet{Rosotti18,Zagaria21}, we impose a zero-flux outer boundary condition at the location at which circumstellar discs are truncated by the outer companion, $r_{\rm t}$. This approach is computationally simpler than by enforcing explicit torques from the outer companion on to the discs \citep[e.g.][]{LinPapaloizou86}, and is valid since the tidal torques exhibit a steep radial dependence, $r^{-4}$, and so their effects quickly become negligible close to the truncation radius \citep{Zagaria21}. Previous works compared the two approaches, and found that they give similar evolutionary results for both gas and dust discs under the influence of an outer companion \citep{Zagaria21}.

Following \citet{Coleman22} we include EUV and X-ray internal photoevaporative winds from the central star (detailed in section \ref{sec:internalPhoto}) as well as winds launched from the outer disc by far ultraviolet (FUV) radiation emanating from nearby massive stars (e.g. O-type stars, see section \ref{sec:externalPhoto}).
We assume that the photoevaporative mass loss rate at any radius in the disc is the maximum of the internally and externally driven rates 
\begin{equation}
    \dot{\Sigma}_{\rm PE}(r) ={\rm max}\left(\dot{\Sigma}_{\rm I,X}(r),\dot{\Sigma}_{\rm E,FUV}(r)\right)
\end{equation}
where the subscripts I and E refer to contributions from internal and external photoevaporation.

We assume that the disc is in thermal equilibrium, and so the temperature is calculated by balancing irradiation heating from the central star, background heating from the residual molecular cloud, viscous heating and blackbody cooling. We neglect any heating or radiation on to the disc from the companion star, since the truncation radius is always less than half the distance between the binary stars, whilst the companion star is assumed to have an evolving protoplanetary disc of it's own that would block much of the radiation from impacting on to the disc. Therefore we assume that the primary star is the main contributor to the irradiation heating of the disc.
To attain thermal equilibrium, we follow \citet[hereafter \citetalias{Coleman21}]{Coleman21} and use an iterative method to solve the following equation \citep{Dangelo12}
\begin{equation}
    Q_{\rm irr} + Q_{\nu} + Q_{\rm cloud} - Q_{\rm cool} = 0
\end{equation}
where $Q_{\rm irr}$ is the radiative heating rate due to the central star, $Q_{\nu}$ is the viscous heating rate per unit area of the disc, $Q_{\rm cloud}$ is the radiative heating due to the residual molecular cloud, and $Q_{\rm cool}$ is the radiative cooling rate.

To take into account the effects of the companion star, we set the outer boundary of our simulation domain to the truncation radius $r_{\rm tr}$. We then treat the evolution of the surface density at the truncation radius by following \citet{Zagaria21} and implementing a zero flux boundary condition $(d(\nu\Sigma r^{1/2}))/dr = 0)$ at $r=r_{\rm tr}$. For equal mass binaries, the truncation radius is approximately 1/3 of the binary separation \citep{Papaloizou77}. This is the location at which the evolution of material in the disc becomes dominated by the tidal torques exerted from the companion star as opposed to the viscosity in the disc. With this balance being dependent on the mass ratio of the stars, it is not always at $\sim1/3$ of the binary separation, and as such may be closer or further away depending on if the star is the primary or secondary star. For determining the truncation radius in this work, we utilise the results found in Table 1 of \citet{Papaloizou77}.

\subsubsection{Internal Photoevaporation}
\label{sec:internalPhoto}
The absorption of high energy radiation from the host star by the disc can heat the gas above the local escape velocity, and drive internal photoevaporative winds. EUV irradiation creates a layer of ionised hydrogen with temperature $\sim$10$^4$~K \citep{Clarke2001}, whilst X-rays penetrate deeper into the disc and are capable of heating the gas up to around $\sim$10$^4$~K \citep{Owen10} and so for low mass stars are expected to generally dominate over the EUV for setting the mass loss rate. FUV radiation penetrates deeper still, creating a neutral layer of dissociated hydrogen with temperature of roughly 1000K \citep{Matsuyama03}. The overall interplay between the EUV, FUV and X-rays is a matter of ongoing debate.  \cite{Owen12} find that including the FUV heating simply causes the flow beneath the sonic surface to adjust, but otherwise retains similar mass loss rates. However other works suggest a more dominant role of the FUV \citep[e.g.][]{Gorti09,Gorti15}. Recent models including all three fields suggest a more complicated interplay \citep[e.g.][]{Wang17, Nakatani18,Ercolano21}. Additionally, the outcome also depends sensitively on how the irradiated spectrum is treated \citep{Sellek22}. 

The radiation hydrodynamic models of \cite{Owen12} used pre-computed X-ray driven temperatures as a function of the ionisation parameter ($\xi = L_X / n /r^2$) wherever the column density to the central stars' radiation $n$ is less than $10^{22}$cm$^{-2}$ (thus optically thin). Here, $L_X$ represents the X-ray luminosity of the central star. More recently this approach has been updated with a series of column-dependent temperature prescriptions \citep{Picogna19,Picogna21,Ercolano21}.

We follow \citet{Picogna21} who further build on the work of \cite{Picogna19} and \cite{Ercolano21} and find that the mass loss profile from internal X-ray irradiation is approximated by
\begin{equation}
\label{eq:sig_dot_xray}
\begin{split}
\dot{\Sigma}_{\rm I,X}(r)=&\ln{(10)}\left(\dfrac{6a\ln(r)^5}{r\ln(10)^6}+\dfrac{5b\ln(r)^4}{r\ln(10)^5}+\dfrac{4c\ln(r)^3}{r\ln(10)^4}\right.\\
&\left.+\dfrac{3d\ln(r)^2}{r\ln(10)^3}+\dfrac{2e\ln(r)}{r\ln(10)^2}+\dfrac{f}{r\ln(10)}\right)\\
&\times\dfrac{\dot{M}_{\rm X}(r)}{2\pi r} \dfrac{\msun}{\au^2 {\rm yr}}
\end{split}
\end{equation}
where
\begin{equation}
\label{eq:m_dot_r_xray}
    \dfrac{\dot{M}_{\rm X}(r)}{\dot{M}_{\rm X}(L_{X})} = 10^{a\log r^6+b\log r^5+c\log r^4+d\log r^3+e\log r^2+f\log r+g}
\end{equation}
where $a=-0.6344$, $b=6.3587$, $c=-26.1445$, $d=56.4477$, $e=-67.7403$, $f=43.9212$, and $g=-13.2316$.
We follow \citet{Komaki23} and apply a simple approximation to the outer regions of the disc where the internal photoevaporation rates drop to zero. The reasons given for this sudden drop is that the wind itself blocks radiation from heating the outer regions of protoplanetary discs. However these do not take into account the effects of when the disc and/or the wind become optically thin and therefore ineffective at blocking the radiation.
The temperature of X-ray irradiated gas varies from $\sim 10^3$--$10^4$ K depending on the distance in the disc \citep[e.g.][]{Owen10}. To be conservative we define the radius at which the internal photoevaporation scheme drops off as the gravitational radius for $10^3$ K gas.
We therefore apply the following approximation at radial distances greater than $r_{\rm rgx}$
\begin{equation}
    \dot{\Sigma}_{\rm I,X,ap} = 37.86\times\dot{\Sigma}_{\rm rgx}\left(\frac{r}{r_{\rm rgx}}\right)^{-1.578}
\end{equation}
where $\dot{\Sigma}_{\rm rgx}$ is equal to eq. \ref{eq:sig_dot_xray} calculated at $r=r_{\rm rgx}$, and
\begin{equation}
    r_{\rm rgx} = \dfrac{GM_*}{c_{\rm s}^2}
\end{equation}
where $c_{\rm s}$ is the sound speed for gas of temperature $T=10^3 K$, and $\mu=2.35$.
In the outer regions of the disc the loss in gas surface density due to internal photoevaporation then becomes
\begin{equation}
    \dot{\Sigma}_{\rm I}(r) = \max(\dot{\Sigma}_{\rm I,X}(r),\dot{\Sigma}_{\rm I,X,ap})
\end{equation}

Following \cite{Ercolano21} the integrated mass-loss rate, dependant on the stellar X-ray luminosity, is given as
\begin{equation}
    \log_{10}\left[\dot{M}_{X}(L_X)\right] = A_{\rm L}\exp\left[\dfrac{(\ln(\log_{10}(L_X))-B_{\rm L})^2}{C_{\rm L}}\right]+D_{\rm L},
\end{equation}
in $\msunyr$, with $A_{\rm L} = -1.947\times10^{17}$, $B_{\rm L} = -1.572\times10^{-4}$, $C_{\rm L} = -0.2866$, and $D_{\rm L} = -6.694$. By including additional cooling effects due to the excitation of O from neutral H, \citet{Sellek24} found that the mass loss rates found in \citet{Picogna21} are overestimated by factor $\sim9$, and as such we apply this correction to the equation above.

\subsubsection{External Photoevaporation}
\label{sec:externalPhoto}

In addition to internal winds driven by irradiation from the host star, winds can also be driven from the outer regions of discs by irradiation from external sources. Massive stars dominate the production of UV photons in stellar clusters and hence dominate the external photoevaporation of discs. External photoevaporation has been shown to play an important role in setting the evolutionary pathway of protoplanetary discs \citep{Coleman22}, their masses \citep{Mann14,Ansdell17,VanTerwisga23}, radii \citep{Eisner18} and lifetimes \citep{Guarcello16,ConchaRamirez19,Sellek20} even in weak UV environments \citep{Haworth17,Coleman24MHD,Coleman25,Anania25}.
We do not include shielding of the protoplanetary discs, i.e. by the nascent molecular cloud, that has been shown to have an effect on the effectiveness of external photoevaporation \citep{Qiao22,Qiao23,Wilhelm23,Qiao26}, but instead will infer it's effects on discs by examining weaker environments. Additionally, we also assume that the systems are moving statically, so do not include the effect of flybys at large distances by massive stars that can dramatically increase the FUV field strength for short periods of time \citep{Coleman25FUV,Coleman25SigOri}.

In our simulations, the mass loss rate due to external photoevaporation is calculated by interpolating over the recently updated \textsc{fried} grid \citep{Haworth23}.
This new grid expands on the original version of \textsc{fried} \citep{Haworth18} in terms of the breadth of parameter space in UV field, stellar mass, disc mass and disc radius. The new grid also provides the option to use different PAH-to-dust ratios, which is important because polycyclic aromatic hydrocarbons (PAH) can provide the main heating mechanism in a photodissociation region (PDR, which is the region at the base of an external photoevaporative wind) and their abundance is uncertain. Finally, the new grid provides the option to control whether or not grain growth has taken place in the disc, which affects the opacity in the wind since only small grains are entrained.

We evaluate the \textsc{fried} mass loss rate at each radius from the outer edge of the disc down to the radius that contains 80$\%$ of the disc mass. We choose this value as 2D hydrodynamical models show that the vast majority of the mass loss from external photoevaporation, comes from the outer 20\% of the disc \citep{Haworth19}. The change in gas surface density is then calculated as
\begin{equation}
    \dot{\Sigma}_{\textrm{ext, FUV}}(r) = G_{\rm sm} \frac{\dot{M}_\textrm{{ext}}(r_{\textrm{\textrm{max}}})}{\pi(r^2_\textrm{{d}} - {r_{\textrm{\textrm{max}}}}^2)+A_{\rm sm}}, 
\end{equation}
where $A_{\rm sm}$ is a smoothing area equal to 
\begin{equation}
A_{\rm sm} = \dfrac{\pi(r_{\rm max}^{22}-(0.1 r_{\rm max})^{22})}{11r_{\rm max}^{20}}
\end{equation}
and $G_{\rm sm}$ is a smoothing function
\begin{equation}
    G_{\rm sm} = \dfrac{r^{20}}{r_{\rm max}^{20}}.
\end{equation}
The outer disc radius $r_{\rm d}$ is calculated as the outermost location where the surface density $\Sigma(r)>10^{-4} \rm gcm^{-2}$, whilst $r_{\rm max}$ is the radial location where the mass-loss rate is at a maximum.

The {\sc fried} grid contains multiple subgrids that vary the PAH-to-dust ratio ($\rm f_{PAH}$) and specify whether or not grain growth has occurred. Following recommendations from previous works \citep{Haworth23,Coleman25}, the combination we use here is $\rm f_{PAH}=1$ (an interstellar medium, ISM-like PAH-to-dust ratio) and assume that grain growth has occurred in the outer disc, depleting it and the wind of small grains which reduces the extinction in the wind and increases the mass loss rate compared to when dust is still ISM-like. This combination of parameters results in PAH-to-gas abundances comparable to our limited observational constraints on that value \citep{Vicente13,Schroetter25}.

\subsection{Planetesimal and Planetary Embryo Formation and Accretion}
We utilise the same models as those found in \citetalias{Coleman21} that begin the simulations with a gas and dust disc, and then form the planetesimals and planetary embryos when the required conditions are met. To do this, we assume that as dust in the disc settles to the midplane, it coagulates into large dust particles and subsequently pebbles\footnote{We assume that 80\% of the total solids is converted into pebbles, that is consistent with that found by planet formation models required to form planets through pebble accretion similar to those observed \citep{Brugger20}.}. These pebbles then begin to drift through the disc towards the central star. Since the settling time-scale depends on the orbital period, this creates a pebble production front that moves outwards in the disc, forming pebbles that then drift inwards \citep{Lambrechts12,Lambrechts14}. As the pebbles drift inwards, we assume that they can collect in short-lived traps due to non-axisymmetric perturbations in the discs \citep{Lenz19}. If sufficient quantities of pebbles are able to collect in those traps such that the local particle density exceeds the Roche density and that the local dust-to-gas ratio exceeds unity, then they may undergo gravitational collapse and form planetesimals \citep{Johansen07,Johansen09}. The size distribution of the planetesimals then follows a power-law plus an exponential decay \citep{Johansen15,Schafer17,Abod19}, and we then assume that the most massive planetesimal to form is a planetary embryo. We briefly discuss this process below, but refer the reader to \citetalias{Coleman21} for more detailed descriptions of the procedures.

\subsubsection{Forming planetesimals and planetary embryos}
It is typically assumed that planetesimals form through the gravitational collapse of smaller dust and pebbles, when the local particle density exceeds the Roche density, which can occur when the dust-to-gas ratio exceeds unity \citep{Johansen07,JohansenYoudin2009}. One way of achieving a dust-to-gas ratio of unity is to concentrate dust and pebbles at pressure bumps in the disc. These pressure bumps have been observed to form in numerous local \citep{Johansen09,Simon12,Dittrich13} and global \citep{SteinackerPapaloizou2002,PapaloizouNelson2003,FromangNelson2006} magnetohydrodynamic (MHD) simulations, and more recently in simulations including non-ideal MHD effects \citep{Bai2014,ZhuStoneBai2014,BethuneLesur2016}. They typically arise from localised magnetic flux concentration and the associated enhancement of magnetic stresses.

Whilst it may appear difficult to form long-lived pressure bumps throughout the entire protoplanetary disc, it would be reasonable to assume that short-lived local pressure bumps could form stochastically throughout the disc, enabling local dust-to-gas ratios to exceed unity for short times only. This approach was recently examined by \citet{Lenz19} where they assumed that pressure bumps formed throughout the protoplanetary disc, creating traps for solids that would then collapse and form planetesimals. We follow the approach of \citet{Lenz19} in forming our planetesimals, and is briefly outlined below.

Whilst we assume that the pebble traps are forming throughout the disc, they do possess a significant lifetime before they dissipate. We take this lifetime to be equal to 100 local orbital periods, with their formation/dissipation assumed to occur almost instantaneously. Given that it takes significant time for all of the pebbles at a location to fall to the midplane and begin drifting towards the centre of the system, we only allow the pebble traps to begin converting the trapped pebbles into planetesimals 100 local orbital periods after the pebble growth front has reached a location $r$. We then assume that a specific percentage of pebbles are trapped in the pebble traps, $\epsilon=10\%$, with the remainder drifting past. This could be the case if the pebble traps are not fully azimuthally encompassing, i.e. vortices, and as such there are areas of the disc azimuthally that do not hinder the inward drift of pebbles.
We therefore define $\Sigma_{\rm trap}$ as:
\begin{equation}
\label{eq:sig_trap}
    \Sigma_{\rm trap} = \dfrac{\epsilon M_{\rm flux} \tau_{100}}{2\pi r l}
\end{equation}
where $\tau_{100}$ is equal to 100 local orbital periods, and $l$ is the length scale over which planetesimal formation occurs within the pebble trap, which we take to be equal to $0.01 H_{\rm gas}$ \citep[][see their eq. (3.40)]{Schreiber18}.

Just because the pebble traps are able to trap pebbles, this doesn't necessarily mean that the planetesimals are able to form there. For planetesimals to form through for example gravitational collapse following the streaming instability, we assume that the local dust-to-gas ratio at the disc midplane has to be equal to or exceed unity, i.e. the dust density has to be greater than or equal to the gas density at the disc midplane \citep{Youdin05}. When the dust-to-gas ratio exceeds unity, the growth rates of the fastest growing modes significantly increase, allowing for particles to concentrate on short time-scales, which could then undergo gravitational collapse.
Following \citet{Youdin07}, we define $H_{\rm peb}$ as
\begin{equation}
    H_{\rm peb} = H_{\rm gas}\sqrt{\dfrac{\alpha}{\Stokes}}
\end{equation}
where $\Stokes$ is the Stokes number and assumed to be equal to,
\begin{equation}
\Stokes = {\rm min}~(\Stokes_{\rm drift}, \Stokes_{\rm frag}),
\end{equation}
where $\Stokes_{\rm drift}$ is the drift-limited Stokes number and $\Stokes_{\rm frag}$ is the fragmentation-limited Stokes number.

Then by equating the pebble density to the gas density, we derive the pebble surface density required for the streaming instability to occur, $\Sigma_{\rm SI}$,
\begin{equation}
\label{eq:sig_SI}
\Sigma_{\rm SI} = \Sigma_{\rm gas}\sqrt{\dfrac{\alpha}{\Stokes}}.
\end{equation}
For gravitational collapse to be able to occur, $\Sigma_{\rm trap} \ge \Sigma_{\rm SI}$.

A further condition for planetesimals to form through gravitational collapse is that the local midplane density of pebbles has to be larger than the Roche density, $\rho_{\rm p} > \rho_{\rm Roche}$, where
\begin{equation}
    \rho_{\rm Roche} = \frac{9}{4\pi}\frac{M_*}{r^3}.
\end{equation}
When this condition is met, the self-gravity of the pebbles is strong enough to overcome the Keplerian shear which leads to the gravitational collapse of the pebbles \citep{GoldreichWard73,Johansen14}. Therefore with this criterion, it means that even when the pebble density is greater than the gas density and stable filaments are effectively formed, those filaments will not produce any planetesimals unless $\rho_{\rm p}>\rho_{\rm Roche}$.
Assuming that the pebble density inside the filaments is proportional to the gas density, we have
\begin{equation}
    \rho_{\rm p,SI} = \epsilon_{\rm R}\rho_{\rm gas}
\end{equation}
with $\rho_{\rm p,SI}$ being the pebble density within a filament produced by the streaming instability,  $\rho_{\rm gas}$ being the gas midplane density, and $\epsilon_{\rm R}$ being an enhancement factor. As the streaming instability concentrates the pebbles into denser and denser filaments, the value of $\epsilon_{\rm R}$ will increase such that $\epsilon_{\rm R} \gg 1$ inside the filaments. Given that it is still unclear of how $\epsilon_{\rm R}$ depends on the Stokes number and the local disc metallicity, as well as possibly other simulation parameters, we assume that $\epsilon_{\rm R}$ to be equal to $10^4$ as within our simulations, the Stokes number typically ranges between 0.05 and 0.2, consistent with previous simulations \citep{Johansen15,Yang17,Carrera21}.

Once these conditions are met, we assume that planetesimals are able to form in the disc through gravitational collapse and that they have a specific size. Numerous works have found that the initial mass function of streaming-derived planetesimals can be roughly fitted by a power-law plus an exponential decay \citep{Johansen15,Schafer17,Abod19}. Recently \citet{Abod19} showed that the initial mass function depends only weakly on the aerodynamic properties of the disc and participating solids, when they concluded the effects of particle self-gravity within their streaming instability simulations. Following \citet{Abod19}, the initial mass of function of planetesimals is,
\begin{equation}
\label{eq:N_mpltml}
N(>M_{\rm pltml}) = C_1 M_{\rm pltml}^{1-p} \exp[-M_{\rm pltml}/M_0]
\end{equation}
where $p \simeq 1.3$, and $C_1$ is the normalisation constant set by the integrated probability equalling unity \citep{Meerschaert12},
\begin{equation}
    C_1 = M_{\rm pltml, min}^{p-1}\exp[{M_{\rm pltml, min}/M_0}], 
\end{equation}
where $M_{\rm pltml, min}$ is the minimum planetesimal mass formed by the streaming instability (taken in this work as being equal to $0.01\times M_0$). The characteristic mass $M_0$ denotes the mass where the initial planetesimal mass function begins to steepen with the exponential part of eq. \ref{eq:N_mpltml} beginning to dominate. Therefore $M_0$ can be treated as a proxy for the planetesimal size and given that the majority of the mass in planetesimals that form through the streaming instability is tied up in the most massive planetesimals, we assume that $M_0$ is the average mass and therefore size of the planetesimals that form in our simulations. Given that the tail of the initial mass function is exponential, it is possible to form planetesimals more massive than $M_0$. In their simulations, \citet{Abod19} find that the most massive planetesimals formed are of the order of the gravitational mass, that is the maximum mass where self-gravity forces are stronger than the tidal shear forces emanating from the particle clumps interactions with the local disc, typically an order of magnitude larger than the characteristic mass. As well as the work of \citet{Abod19}, other works have also found similar results regarding the initial planetesimal mass function and expressions for calculating the characteristic mass $M_0$ \citep{Johansen15,Schafer17,Simon16,Li19}. More recently, \citet{Liu20} extrapolated on the simulations from those works, and derived an expression for the characteristic planetesimal mass,
\begin{equation}
\label{eq:MG_Liu}
M_0 = 5\times 10^{-5}\left(\frac{Z}{0.02}\right)^{1/2}\left(\pi\gamma\right)^{3/2}\left(\frac{h}{0.05}\right)^3\left(\frac{M_*}{1\msun}\right)\me,
\end{equation}
where $Z=\Sigma_{\rm peb}/\Sigma_{\rm gas}$ is the local disc metallicity and $\gamma=4\pi G\rho_{\rm gas}/\Omega$ is a self-gravity parameter. Here we take $\Sigma_{\rm peb}$ to be equal to $\Sigma_{\rm SI}$, that is the surface density of pebbles required for the pebble cloud to undergo gravitational collapse. We use $\Sigma_{\rm SI}$ instead of $\Sigma_{\rm trap}$ as we assume that once the requisite mass in pebbles has become trapped, i.e. $\Sigma_{\rm SI}$, then the pebble cloud undergoes gravitational collapse. This is valid, so long as the pebble midplane density is larger than the Roche density, which is the case when $\Sigma_{\rm SI}>\Sigma_{\rm trap}$, except for the innermost region of the disc, well inside in the iceline.

When the planetesimals form through gravitational collapse, numerous works found relations for the planetesimal initial mass functions \citep{Johansen15,Schafer17,Abod19}. Whilst the number of massive planetesimals drops exponentially at masses greater than the characteristic mass, should there be enough mass being converted from pebbles to planetesimals, then a number of more massive planetesimals will be able to form. It is these more massive planetesimals that we assume to be the planetary embryos that will eventually grow into the planets that are observed in planetary systems. In our simulations, we assume that for each planetesimal formation event, the largest planetesimal that forms is a planetary embryo. Using eq. \ref{eq:N_mpltml} we calculate the largest single body that forms in each event, (i.e. $N(>M_{\rm emb})=1$). The masses of these planetary embryos is typically an order of magnitude larger than the neighbouring planetesimals \citepalias{Coleman21}.

\subsubsection{Pebble Accretion}
Since the pebbles in our simulations are contributing to the planetesimal formation process, the amount of pebbles drifting past the planetary embryos is much reduced compared to previous works \citep[e.g.][]{Lambrechts14,Bitsch15,Coleman19}. Still, as the remaining pebbles drift through the disc, they are able to be accreted more efficiently than planetesimals when passing through a planetary embryo's Hill or Bondi sphere. This is due to the increased gas drag forces that allowed them to become captured by the planetary embryo's gravity \citep{Lambrechts12}. To calculate this accretion rate, we follow \citet{Johansen17} by distinguishing between the Bondi regime (small bodies) and the Hill regime (massive bodies). The Bondi accretion regime occurs for low mass bodies where they do not accrete all of the pebbles that pass through their Hill sphere, i.e. the body's Bondi radius is smaller than the Hill radius. Once the Bondi radius becomes comparable to the Hill radius, the accretion rate becomes Hill sphere limited, and so the body accretes in the Hill regime. Within our simulations, planetary embryos typically begin accreting in the Bondi regime before transitioning to the Hill regime when they reach the transition mass,
\begin{equation}
\label{eq:mtrans}
    M_{\rm trans} \propto \eta^3 M_*.
\end{equation}
A further distinction within the two regimes, is whether the body is accreting in a 2D or a 3D mode. This is dependent on the relation between the Hill radius of the body and the scale height of the pebbles in the disc. For bodies with a Hill radius smaller than the scale height of pebbles, the accretion is in the 3D mode since pebbles are passing through the entire Hill sphere, whilst for bodies with a Hill radius larger than the pebble scale height, regions of the Hill sphere remain empty of pebbles and as such the accretion rate becomes 2D as the body's mass increases. Following \citet{Johansen17} the equations for the 2D and 3D accretion rates are
\begin{equation}
\dot{M}_{\rm 2D} = 2 R_{\rm acc} \Sigma_{\rm peb} \delta v,
\end{equation}
and
\begin{equation}
\dot{M}_{\rm 3D} = \pi R_{\rm acc}^2 \rho_{\rm peb} \delta v,
\end{equation}
where $\Sigma_{\rm peb}$ is the pebble surface density, while $\rho_{\rm peb}$ is the midplane pebble density. Here $\delta v = \Delta v + \Omega R_{\rm acc}$ is the approach speed, with $\Delta v$ being the sub-Keplerian velocity. The accretion radius $R_{\rm acc}$ depends on whether the accreting object is in the Hill or Bondi regime, and also on the friction time of the pebbles. In order for pebbles to be accreted they must be able to significantly change direction on time-scales shorter than the friction time. This inputs a dependence of the friction time onto the accretion radius, forming a criterion accretion radius $\hat{R}_{\rm acc}$ which is equal to 
\begin{equation}
 \hat{R}_{\rm acc} = \left( \frac{4 t_{\rm f}}{t_{\rm B}} \right)^{1/2} R_{\rm B},
\end{equation}
for the Bondi regime, and:
\begin{equation}
 \hat{R}_{\rm acc} = \left(  \frac{\Omega t_{\rm f}}{0.1} \right)^{1/3} R_{\rm H},
 \label{Racc_Hill}
\end{equation}
for the Hill regime.
Here $R_{\rm B}$ is the Bondi radius, while $R_{\rm H}$ is the Hill radius, $t_{\rm B}$ is the Bondi sphere crossing time, and $t_{\rm f}$ is the friction time.
The accretion radius is then equal to
\begin{equation}
    R_{\rm acc} = \hat{R}_{\rm acc} \exp[-\chi(t_{\rm f}/t_{\rm p})^{\gamma}]
\end{equation}
where $t_{\rm p} = G M/(\Delta v + \Omega R_{H})^3$ is the characteristic passing time-scale, $\chi = 0.4$ and $\gamma = 0.65$ \citep{OrmelKlahr2010}.

The object then grows by accreting pebbles until it reaches the so-called pebble isolation mass, that is the mass required to perturb the gas pressure gradient in the disc: i.e. the gas velocity becomes super-Keplerian in a narrow ring outside the planet's orbit reversing the action of the gas drag. The pebbles are therefore pushed outwards rather than inwards and accumulate at the outer edge of this ring stopping the core from accreting solids \citep{PaardekooperMellema06,Rice06}. Initial works found that the pebble isolation mass was proportional to the cube of the local gas aspect ratio \citep{Lambrechts14}. More recent work however has examined what effects disc viscosity and the stokes number of the pebbles have on the pebble isolation mass, finding that small pebbles that are well coupled to the gas are able to drift past the pressure bump exterior to the planet's orbit \citep{Ataiee18,Bitsch18}. To account for the pebble isolation mass whilst including the effects of turbulence and stokes number, we follow \citet{Bitsch18}, and define a pebble isolation mass-to-star ratio,
\begin{equation}
q_{\rm iso} = \left(q_{\rm iso}^{\dagger} + \frac{\Pi_{\rm crit}}{\lambda} \right) \frac{M_{\oplus}}{M_*}
\end{equation}
where $q_{\rm iso}^\dagger  = 25 f_{\rm fit}$, and
\begin{equation}
\label{eq:ffit}
 f_{\rm fit} = \left[\frac{H/r}{0.05}\right]^3 \left[ 0.34 \left(\frac{\log(\alpha_3)}{\log(\alpha)}\right)^4 + 0.66 \right] \left[1-\frac{\frac{\partial\ln P}{\partial\ln r } +2.5}{6} \right] \ ,
\end{equation}
with $\alpha_3 = 0.001$.

\subsubsection{Planetesimal Accretion}
Since we are counting planetary embryos as single bodies in the disc, we do not class planetesimals as super particles such as seen in previous works \citep[e.g.][]{ColemanNelson14,ColemanNelson16,ColemanNelson16b}. Instead we follow other works that treat the accretion of planetesimals from an evolving surface density, essentially treating the planetesimals as a fluid-like disc \citep[e.g.][]{Alibert2006,Ida13,Mordasini15}. We follow \citet{Fortier13} in calculating the planetesimal accretion rate that depends on the inclination and eccentricity evolution of the planetesimals. The planetesimal surface density $\Sigma_{\rm pltml}$ evolves as planetesimals are accreted by planetary embryos. We also evolve the eccentricity rms $e_{\rm pltml}$ and inclination rms $i_{\rm pltml}$ by solving the differential equations for self-stirring \citep{Ohtsuki99}, gravitational stirring by nearby planetary embryos \citep{Ohtsuki99} and also the effects of gas disc damping \citep{Adachi,Inaba01,Rafikov04}.

The planetesimal accretion rate is equal to
\begin{equation}
\label{eq:mdotpltml}
    \dot{M}_{\rm pltml} = \Omega \bar{\Sigma}_{\rm pltml} R_{\rm H}^2 P_{\rm coll}
\end{equation}
where $\bar{\Sigma}_{\rm pltml}$ is the average planetesimal surface density in the planetary embryos' feeding zone (taken here to be equal to 10 Hill radii) and $P_{\rm coll}$ is the collision probability following \citet{Inaba01},
\begin{equation}
\label{eq:P_col}
P_{\rm coll}=\min\left( {P}_{\rm med},\left({P}^{-2}_{\rm high}+{P}^{-2}_{\rm low}\right)^{-1/2}\right)
\end{equation}
where the individual components are equal to
\begin{equation}
    P_{\rm high} = \frac{(r_{\rm emb}+r_{\rm pltml})^2}{2\pi R_{\rm H}^2}\left(I_{\rm F}(\beta)+\frac{6R_{\rm H}I_{\rm G}(\beta)}{(r_{\rm emb}+r_{\rm pltml})\tilde{e}^2} \right)
\end{equation}
\begin{equation}
    P_{\rm med} = \frac{(r_{\rm emb}+r_{\rm pltml})^2}{4\pi R_{\rm H}^2\tilde{i}}\left(17.3+\frac{232R_{\rm H}}{r_{\rm emb}+r_{\rm pltml}}\right)
\end{equation}
\begin{equation}
    P_{\rm low} = 11.3\left(\frac{r_{\rm emb}+r_{\rm pltml}}{R_{\rm H}}\right)^{1/2}
\end{equation}
where $P_{\rm high}$, $P_{\rm med}$ and $P_{\rm low}$ are the collision probabilities for different velocity regimes that depend on the random velocities of the planetesimals. The quantities $\tilde{e}$ and $\tilde{i}$ are the reduced eccentricities ($\tilde{e}=ae/R_{\rm H}$) and inclinations ($\tilde{i}=ai/R_{\rm H}$). These quantities indicate which velocity regime the planetesimals are found in depending on their relative velocities, with: the high-velocity regime for $\tilde{e},\tilde{i} \ge 2$, the medium velocity regime being for $2\ge\tilde{e},\tilde{i} \ge 0.2$ and the low-velocity regime for $\tilde{e},\tilde{i} < 0.2$. The variable $\beta$ in the above equations is equal to $\tilde{i}/\tilde{e}$ and the functions $I_{\rm F}(\beta)$  $I_{\rm G}(\beta)$ are well approximated by
\begin{equation}
    I_{\rm F}(\beta) \backsimeq \frac{1+0.95925\beta+0.77251\beta^2}{\beta(0.13142+0.12295\beta)}
\end{equation}
\begin{equation}
    I_{\rm F}(\beta) \backsimeq \frac{1+0.3996\beta}{\beta(0.0369+0.048333\beta+0.006874\beta^2)}
\end{equation}
for $0\le\beta\le 1$, which is the range of $\beta$ values within this work \citep{Chambers06}.

Since the planetesimal eccentricities and inclinations have a large impact on the planetesimal accretion rate, it is necessary to define these values and allow them to evolve over the course of the disc lifetime. Planetesimals experience gas drag from the disc which acts to damp the eccentricities whilst simultaneously experiencing gravitational interactions with planetary embryos as well as gravitational interactions and minor collisions with fellow planetesimals. In incorporating these processes into the evolution of the planetesimal eccentricities and inclinations, we obtain
\begin{equation}
    \dfrac{de^2}{dt}=\dfrac{de^2}{dt}\bigg|_{\rm drag}+ \dfrac{de^2}{dt}\bigg|_{\rm pltml} + \dfrac{de^2}{dt}\bigg|_{\rm emb}
\end{equation}
\begin{equation}
    \dfrac{di^2}{dt}=\dfrac{di^2}{dt}\bigg|_{\rm drag}+ \dfrac{di^2}{dt}\bigg|_{\rm pltml} + \dfrac{di^2}{dt}\bigg|_{\rm emb}
\end{equation}
where the subscripts `drag', `pltml' and `emb' refer to the contributions from gas drag damping, mutual stirring by planetesimals, and gravitational stirring by planetary embryos. For the calculation of the three terms affecting the eccentricity and inclination evolution, we follow \citet{Fortier13} where the equations and contributions of these terms can be found in their equations 31--53. With the eccentricity and inclination now known for the planetesimals, we allow the planetary embryos to accrete planetesimals using eq. \ref{eq:mdotpltml}, and remove the accreted mass from the planetary embryos local planetesimal surface densities.

Similar to the treatment of the evolution of the gas disc, we assume that the binary companion only acts to truncate the disc at the truncation radius and will therefore have negligible direct effects on the planetesimal and embryo formation processes described above that occur deep within protoplanetary discs. This is again due the rapid fall off of the tidal torque interior to the truncation radius that acts to perturb the disc and increase the eccentricities of planetesimals and planetary embryos. Since the majority of the planetesimals and planetary embryos form well inside the disc, away from the outer edge, the tidal torques from the outer companion should have little direct effect on their formation and evolution. Instead we focus on the effects that the truncation of the gas disc due to these perturbations, will have on the formation and evolution of the planetesimals and planetary embryos.

\subsection{Simulation Parameters}
\label{sec:pop_parameters}
In this work, we aim to explore the simultaneous effects of the external radiation environment and the presence of a wide binary on the formation of planetesimals and planetary embryos. We will base our studies around Solar mass stars, but will later explore how these effects vary across stellar mass.

For the presence of a wide binary star, it is already known that they truncate protoplanetary discs to between 1/4--1/2$\times$ the separation of the binary \citep{Papaloizou77}. For equal mass stars, it is found to be equal to $\sim1/3$ of the binary separation. Instead of varying the separation of potential binary stars, we modify the outer truncation radius of the discs that their evolution can be representative of multiple outer binary masses at varying separations\footnote{Note that we assume the binaries are on circular coplanar orbits.}. We therefore explore truncation radii $r_{\rm t}$ between 5--250$\au$. For the external photoevaporative mass-loss rates, we vary the strength of the local environment, ranging from 10 $\rm G_0$ to $10^5 \rm G_0$, appropriate for most star-forming regions, e.g. Taurus or Orion \citep[see][for a review]{PastPresentFuture2025}.

We initialise our discs following \citet{Lynden-BellPringle1974}
\begin{equation}
    \Sigma = \Sigma_0\left(\frac{r}{1\au}\right)^{-1}\exp{\left(-\frac{r}{r_{\rm C}}\right)}
\end{equation}
where $\Sigma_0$ is the normalisation constant set by the total disc mass, and $r_{\rm C}$ is the scale radius, effectively representing the initial compactness of the disc which we set to 50$\au$.
\citet{Haworth20} found that the maximum mass a disc could be before it went gravitationally unstable was equal to
\begin{equation}
    \label{eq:max_disc_mass}
    \dfrac{M_{\rm d, max}}{M_*} < 0.17 \left(\dfrac{r_{\rm ini}}{100\au}\right)^{1/2}\left(\dfrac{M_*}{\msun}\right)^{-1/2}
\end{equation}
where $r_{\rm ini}$ is the initial disc radius, and taken to be equal to the minimum of the truncation radius or 100$\au$. We set the initial disc mass to being equal to $0.5 \times M_{\rm d, max}$, and expect the effects of the disc mass on the formed planetesimal and embryo masses to be the same as that found in \citetalias{Coleman21}, where the formed masses scaled roughly linearly with the disc mass. Depending on the truncation radius, these correspond to disc masses ranging between 0.019--0.083$\msun$. As well as the main parameters above, we also explore the effects of varying the viscous $\alpha$ parameter within our simulations. We explore $\alpha$ values of $10^{-4}$, $10^{-3.5}$, and $10^{-3}$. These values are consistent with those found from theoretical \citep{Arlt04,Nelson13,Latter16,Flock17,Flock20,Lesur23,Coleman24}, as well as observational studies \citep{Ansdell18,Sierra19,Rosotti20,Trapman20,Coleman26d203}. Table \ref{tab:params} shows the simulation parameters that we used in this work.

\begin{table}
    \centering
    \begin{tabular}{c|c}
    \hline
   Fixed Simulation Parameters & Value \\
    \hline
        $r_{\rm in} (\au)$ & 0.04 \\
        $r_{\rm out} (\au)$ & 500 \\
        $M_{*} (\msun)$ & 1 \\
        $T_* (K)$ & 4300 \\
        $R_* (\rsun)$ & 2 \\
        $L_X (\log_{10}(\rm erg~s^{-1}))$ & 30.5 \\
        $r_{\rm C} (\au)$ & 50 \\
        Z & 0.01 \\
        \hline
        Variable Simulation Parameters & Value \\
        \hline
        $r_{\rm t} (\au)$ & 5--250\\
        $M_{\rm d} (\msun)$ & 0.019--0.085 \\
        $\alpha$ & $10^{-4}$, $10^{-3.5}$, $10^{-3}$ \\
        UV field ($\rm G_0$) & 10--$10^5$ \\
        
    \hline
    \end{tabular}
    \caption{Simulation Parameters for the models.}
    \label{tab:params}
\end{table}

\section{Example Disc Evolution}
\label{sec:disc_evol}

\begin{figure*}
\centering
\includegraphics[scale=0.4]{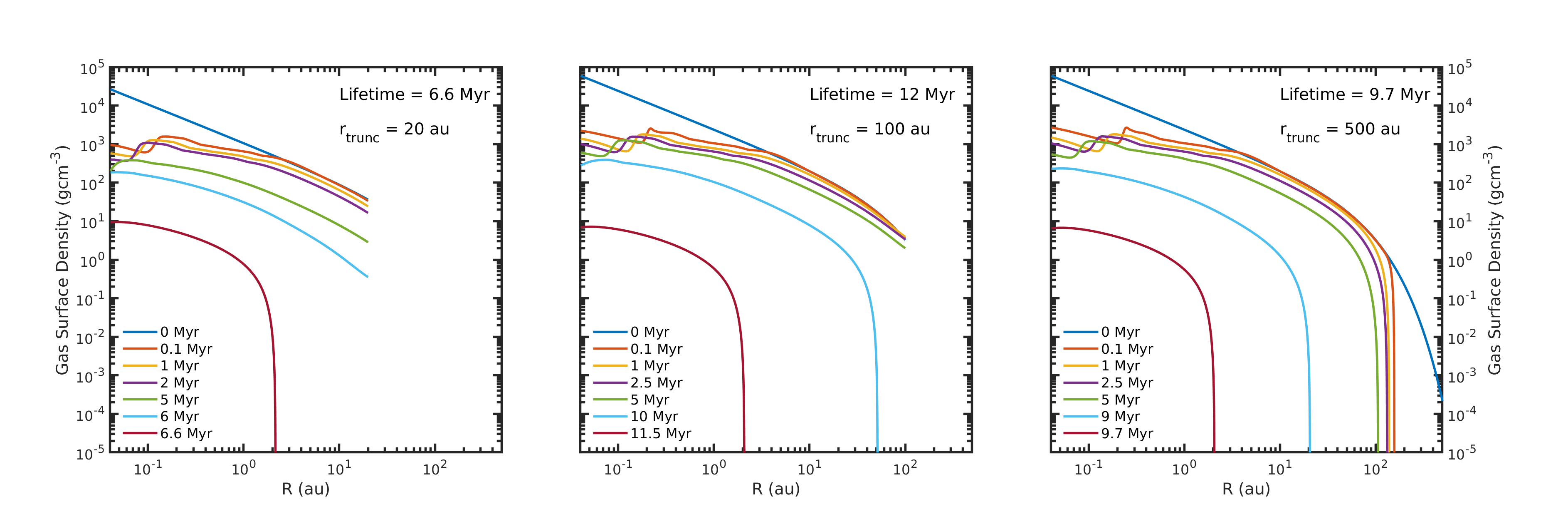}
\caption{Temporal evolution of gas surface densities for viscous discs truncated by an outer companion at distances of $20 \au$ (left-hand panel), $100 \au$ (middle panel), and $500 \au$ (right-hand panel). All discs evolved in a $100 \rm G_0$ environment with the viscous parameter $\alpha=10^{-3}$. Note that the line colours correspond to different simulation times in each panel (with the respective times shown in the panel legends), due to the differences in disc lifetimes.}
\label{fig:disc_trunc}
\end{figure*}

We now explore the effects of the truncation and the external radiation field on the evolution of protoplanetary discs. Focusing initially on the effects of truncation by an outer companion, Fig. \ref{fig:disc_trunc} shows the evolution of the gas surface density for three protoplanetary discs where the truncation radius was set to $20\au$ (left-hand panel), $100\au$ (middle panel), and $500\au$ (right-hand panel). The extended case with $r_{\rm trunc}=500\au$ is effectively equivalent to the evolution of protoplanetary discs around single stars \citep[see for example][]{Coleman25}. We set $\alpha=10^{-3}$ and the external UV field to $100 \rm G_0$.
Given that the truncation of the circumstellar discs by the secondary star is expected to occur on only a few hundred orbital time-scales \citep{Kley08}, we assume that this has already occurred at the start of the simulations.
In all of the discs in Fig. \ref{fig:disc_trunc}, they evolved from their initial profile shown by the dark blue lines in the upper right of each panel. As they evolved over time, the discs lost mass through accretion on to the central star, and through photoevaporative winds, resulting in the surface density profiles moving inward over time. This resulted in the discs being truncated by external photoevaporation, as well as by that expected from the companion star. This truncation by external photoevaporation is especially seen in the right-hand panel of Fig. \ref{fig:disc_trunc} where the disc is truncated down to $150\au$ after $\sim 0.1$ Myr. For the discs with smaller truncation radii due to the companion, they initially evolved quicker over time since those discs were initially less massive and extended. This is similar to what was found in previous works exploring the effects of truncation from an outer companion on discs only evolving through viscosity and internally driven photoevaporative winds \citep{Rosotti18}, where more compact discs evolved faster, mainly due to internal photoevaporative winds operating effectively over a larger region of the disc. For the disc around the extremely wide binary in the right-hand panel of Fig. \ref{fig:disc_trunc}, it had a shorter lifetime than that shown in the middle panel where the disc was truncated to $100 \au$ by the companion star, even though the initial disc masses and profiles were similar. This is due to viscous movement of gas being able to resupply the external photoevaporative winds, reducing the long-term supply of gas to the intermediate regions of the disc between 10--100$\au$, which allowed the disc to evolve more quickly at later times.

\begin{figure*}
\centering
\includegraphics[scale=0.4]{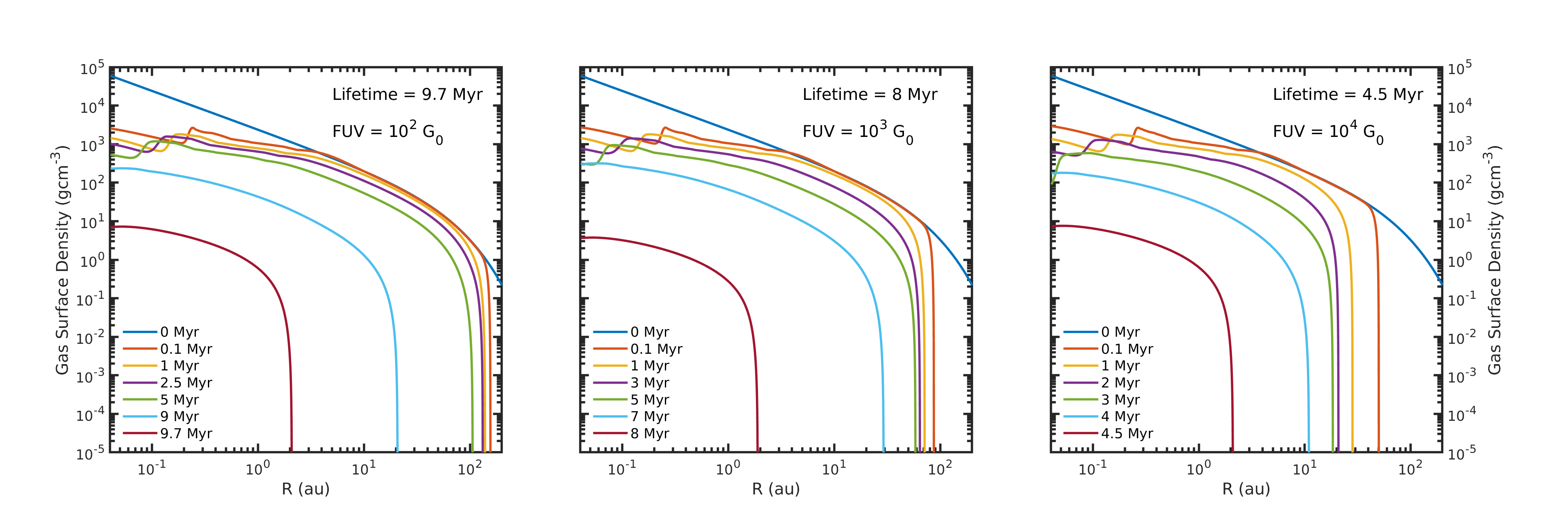}
\caption{Temporal evolution of gas surface densities for viscous discs in different UV environments of $100 \rm G_0$ (left-hand panel), $10^3 \rm G_0$ (middle panel), and $10^4 \rm G_0$. The discs were all initially truncated by an outer companion to $200 \au$, whilst $\alpha = 10^{-3}$. Note that the line colours correspond to different simulation times in each panel (with the respective times shown in the panel legends), due to the differences in disc lifetimes.}
\label{fig:disc_g0}
\end{figure*}

With the effects of truncation by the wide binary acting as expected from previous works \citep[e.g.][]{Rosotti18}, we now show how the local radiation environment affects the evolution of discs that are already truncated by an outer companion. Figure \ref{fig:disc_g0} shows the evolution of three different discs, initially truncated to $200\au$ by the outer companion, whilst being placed in different strength UV environments of $100 \rm G_0$ (left-hand panel), $10^3 \rm G_0$ (middle panel) and $10^4 \rm G_0$ (right-hand panel). Similar to previous works \citep[e.g.][]{Coleman22,Coleman25}, the effects of external photoevaporation increases as the UV field strength increases, that can act to quickly truncate discs down from their initial values. Indeed, the disc in the $100 \rm G_0$ field (left-hand panel) only truncated down to $\sim 150\au$ after 0.1 Myr, whilst the disc in the $10^4 \rm G_0$ field was truncated down to just $50 \au$ over the same time frame. This rapid truncation of discs in the strong UV environments, implies that the effects of truncation by the binary companion may be less important there.

\begin{figure}
\centering
\includegraphics[scale=0.4]{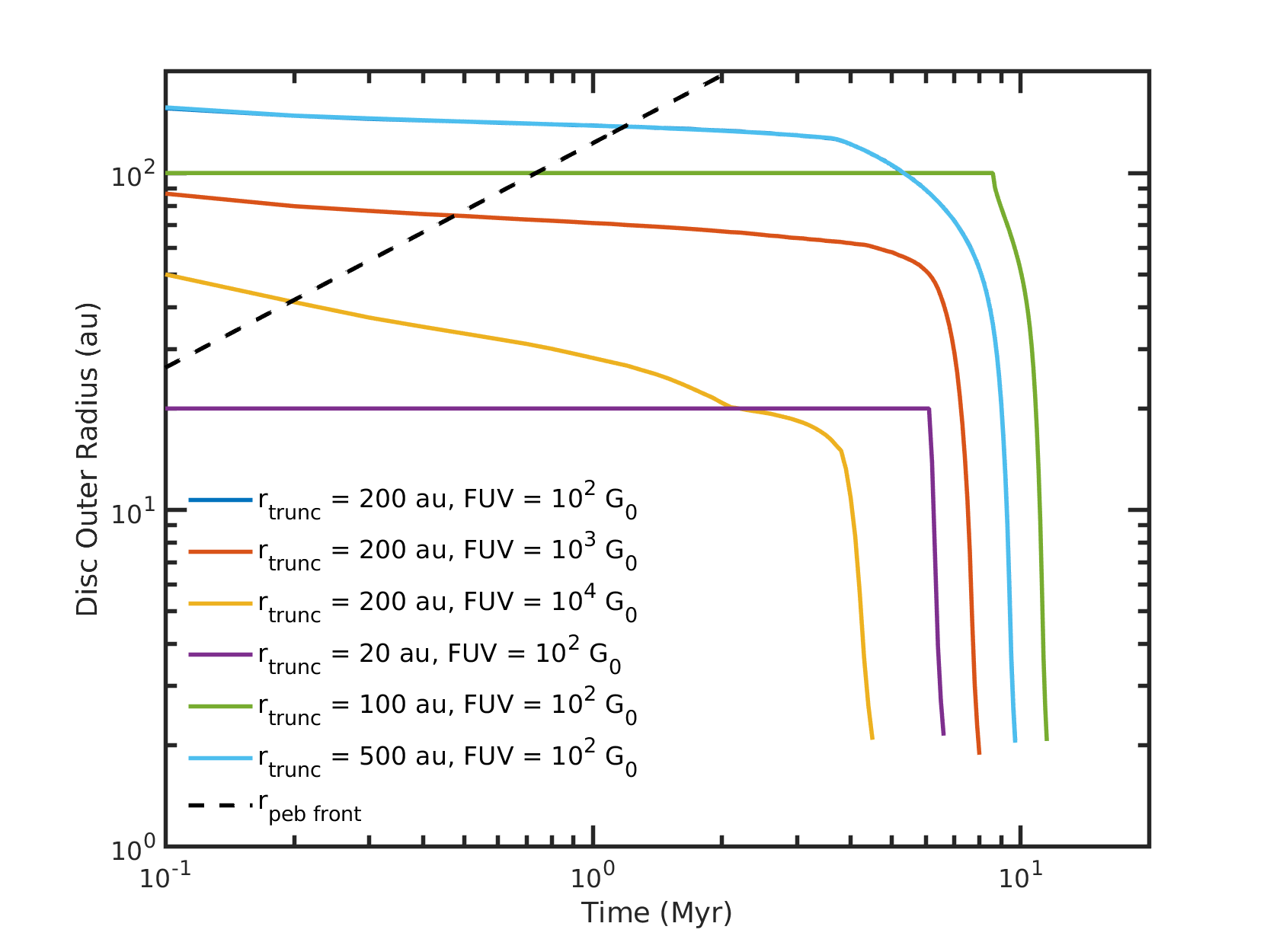}
\caption{Temporal evolution of the disc outer radius for all of the discs shown in Figs. \ref{fig:disc_trunc} and \ref{fig:disc_g0}. The black dashed line shows the pebble production front over time.}
\label{fig:disc_router}
\end{figure}

\section{Role of the pebble cutoff time}
\label{sec:peb_cutoff}

With the section above illustrating how the evolution of protoplanetary discs is affected by both external photoevaporation, and truncation from an outer companion, we now explore the effects on the growth of planets and planetesimals. As has been shown in previous works, in the pebble accretion paradigm where planetary growth is dominated by pebble accretion, increasing levels of disc truncation due to external photoevaporation acts to reduce the effectiveness at which planets form \citep{Winter22,Qiao23,Qiao26}. Indeed, \citet{Qiao23} showcased this effect for single planets forming in protoplanetary discs experiencing varying levels of UV radiation, where those planets in weak UV environments were able to reach super-Earth to Neptune masses, whilst those in extremely strong UV environments remained at sub-terrestrial masses. The main cause of the lack of growth for planets in the strong environments was found to be due to the pebble production front, where dust settles in the disc and coagulates into pebbles before drifting inwards, that reaches the disc outer edge before planets are able to accrete significant quantities of pebbles and substantially increase their mass. Once at the disc outer edge, the pebble production front ceases to form pebbles, thus cutting off the supply of pebbles to planets. This is termed the pebble cutoff time. Since this depends on the speed of truncation of the outer disc edge, it can be expected that the presence of the outer companion may also have similar effects to that of external photoevaporation.

To explore the evolution of the disc outer edge, Fig. \ref{fig:disc_router} shows how the outer edges of the discs presented in Fig. \ref{fig:disc_trunc} and \ref{fig:disc_g0} evolve over time. The dashed black line denotes the evolution of the pebble production front. Looking at those discs where the truncation due to companion was set to 200 $\au$ and the strength of the UV field was varied (blue - $10^2 \rm G_0$, red - $10^3 \rm G_0$, yellow - $10^4 \rm G_0$), it is clear that the disc outer radius is sensitive to the strength of the radiation field. Indeed, in the strongest UV field, the yellow line shows the disc outer radius quickly dropping to 50 $\au$ after only 0.1 Myr, whilst the disc in the weak UV field was still extended to $\sim 150\au$. What is additionally clear is that as the UV field strength increases the time at which the pebble production front crosses the disc outer edge occurs earlier in the disc lifetime, resulting in pebble accretion cutting off earlier in the disc lifetime. Similarly, when determining the effects of the different truncation radii due to the outer companion (shown by the blue - 200 $\au$, green - 100 $\au$, and purple - 20 $\au$ lines), it is clear that the pebble production front reaches the disc outer edge earlier for smaller truncation radii, as would be expected for more compact discs.

\begin{figure}
\centering
\includegraphics[scale=0.5]{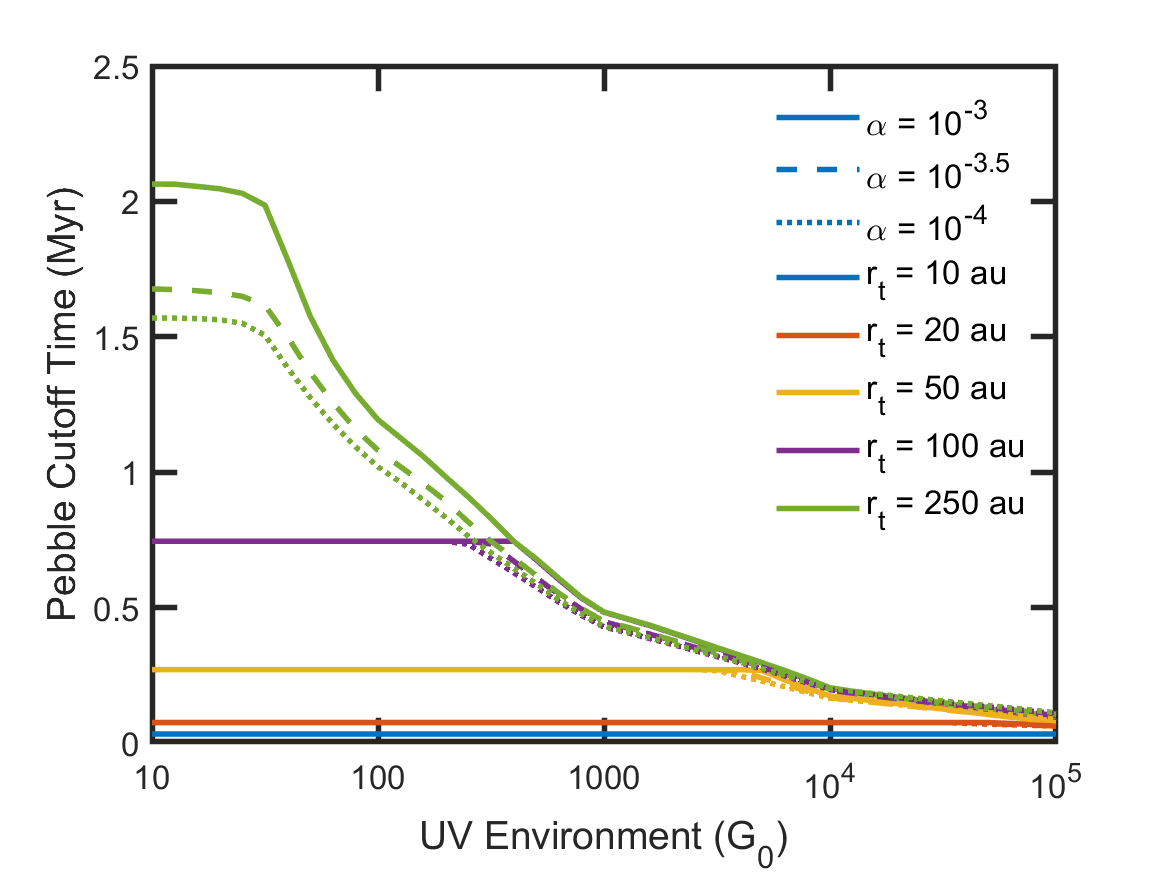}
\caption{Pebble cutoff time as a function of the UV environment. The truncation radius due to the outer companion is highlighted by the different colours showing $r_{\rm t}=10\au$ (blue), $r_{\rm t}=20\au$ (red), $r_{\rm t}=50\au$ (yellow), $r_{\rm t}=100\au$ (purple) and $r_{\rm t}=250\au$ (green). The strength of the viscous turbulent parameter $\alpha$ is shown by the different line styles with $\alpha=10^{-3}$ shown by solid lines, $\alpha=10^{-3.5}$ by dashed lines, and $\alpha=10^{-4}$ by the dotted lines.}
\label{fig:peb_cutoff_g0}
\end{figure}

Looking further into the effects of external photoevaporation on the pebble cutoff time, that is the time when the pebble production front reaches the disc outer edge, Fig. \ref{fig:peb_cutoff_g0} shows the pebble cutoff time as a function of the strength of the UV environment. The line colours denote discs with different initial truncation radii due to the outer companion, whilst the different line styles represent different levels of the viscous $\alpha$ parameter. Looking initially at the green lines, showing discs with initial truncation radii of 250 $\au$, it is clear that in the weakest environments, $\leq40 \rm G_0$, the pebble cutoff time was similar at around 2.1 Myr since even weak levels of photoevaporation are able to slightly truncate protoplanetary discs \citep{Coleman24MHD,Anania25}. As the levels of UV irradiation increased, the pebble cutoff time can be seen to decrease, reaching $\sim1.2$ Myr for $10^2 \rm G_0$, 0.5 Myr for $10^3 \rm G_0$, and 0.2 Myr for $10^4 \rm G_0$. Whilst for discs with large binary separations inducing large truncation radii the pebble cutoff time was mainly set by external photoevaporation, this was less the case for discs with smaller binary truncation radii in weak UV environments. Indeed, the purple line in Fig. \ref{fig:peb_cutoff_g0} indicating the pebble cutoff time for discs with truncation radii of 100 $\au$, shows that the pebble cutoff was $\sim0.75$ Myr for all discs in UV environments up to a strength of $\sim 400 \rm G_0$, with the cutoff time being dictated by the binary truncation radius. As the UV environment increased in strength, external photoevaporation took over as the dominant force in determining when the pebble supply was cut off in the disc for ever decreasing binary truncation radii.

\begin{figure}
\centering
\includegraphics[scale=0.5]{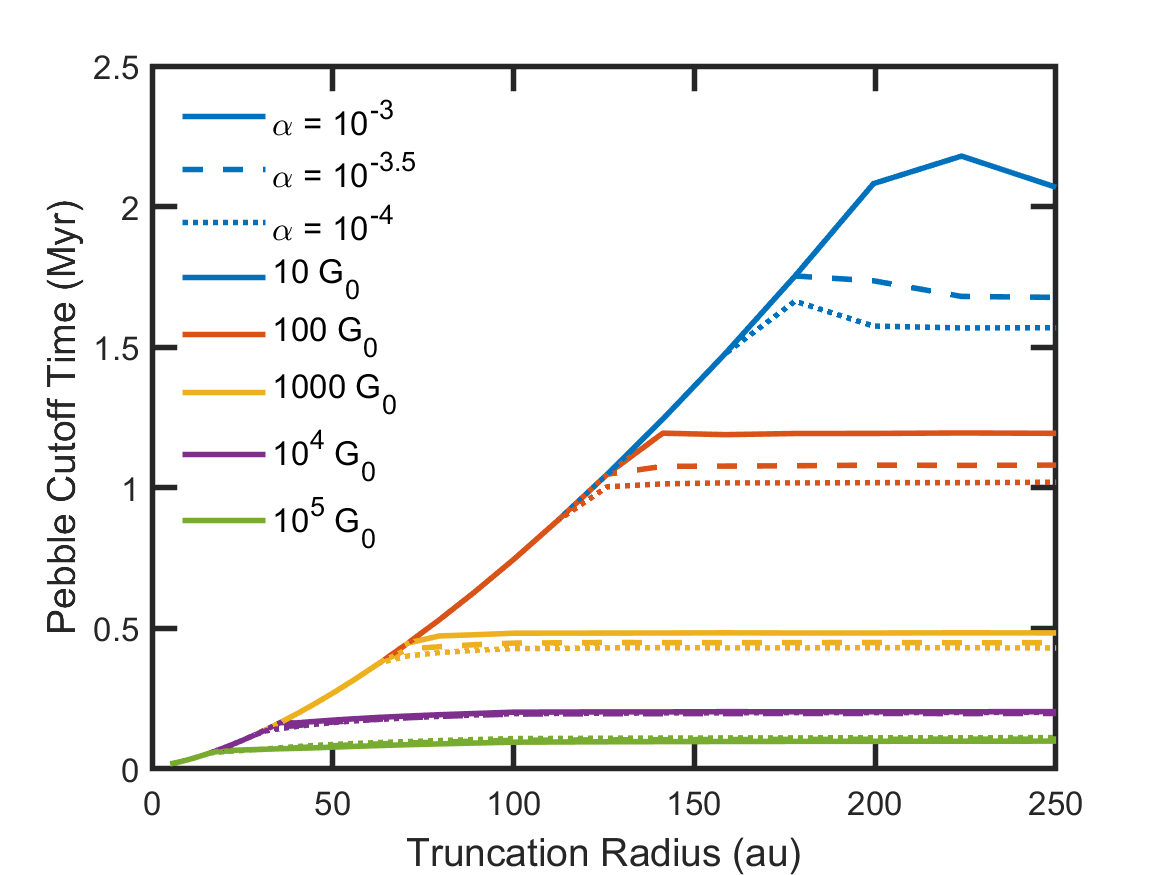}
\caption{Pebble cutoff time as a function of the truncation radius due to the outer companion. Diferent UV field strengths are highlighted by the different colours showing $10\rm G_0$ (blue), $10^2\rm G_0$ (red), $10^3\rm G_0$ (yellow), $10^4\rm G_0$ (purple) and $10^5\rm G_0$ (green). The strength of the viscous turbulent parameter $\alpha$ is shown by the different line styles with $\alpha=10^{-3}$ shown by solid lines, $\alpha=10^{-3.5}$ by dashed lines, and $\alpha=10^{-4}$ by the dotted lines..}
\label{fig:peb_cutoff_trunc}
\end{figure}

Similar to Fig. \ref{fig:peb_cutoff_g0}, Fig. \ref{fig:peb_cutoff_trunc} shows the pebble cutoff time as a function of the binary truncation radius. The colours now show the strength of the external UV field ranging from 10 $\rm G_0$ (blue) to $10^5~\rm G_0$ (green). The line styles again show the different viscous $\alpha$ parameters. The effects of the external radiation field are clear here now, with those discs evolving in strong UV environments, $\geq 10^4\rm G_0$, having their pebble cutoff times being similar for nearly all values of the truncation radius. Only those discs with extremely small truncation radii, were the pebble cutoff times determined by the binary companion. As the binary truncation radius increased, it began to dominate over the external radiation field in determining the pebble cutoff time. This can be seen as the different coloured lines level off at larger pebble cutoff times for different UV environments, with the truncation radius dictating the pebble cutoff time before then. Interestingly, the effects of the viscous $\alpha$ parameter on the pebble cutoff time can only really be found for the more extended discs, and those in weak UV environments, with larger $\alpha$ values extending the time at which the pebble production front reaches the disc outer edge. This is due to the viscous expansion of the gas being able to maintain a larger disc radius by balancing against the external photoevaporation rate. For stronger UV environments, external photoevaporation sets the equilibrium radii, where viscous expansion matches the external photoevaporative mass loss rate, at similar locations of the disc, irrespective of the strength of $\alpha$. Similarly for smaller discs, due to smaller truncation radii, the effects of $\alpha$ are also minimally seen there, since the viscous expansion for all of the $\alpha$ values studied here is able to expand to the truncation radius due to the binary, or be shrunk to similar radii by strong UV fields.

\begin{figure}
\centering
\includegraphics[scale=0.5]{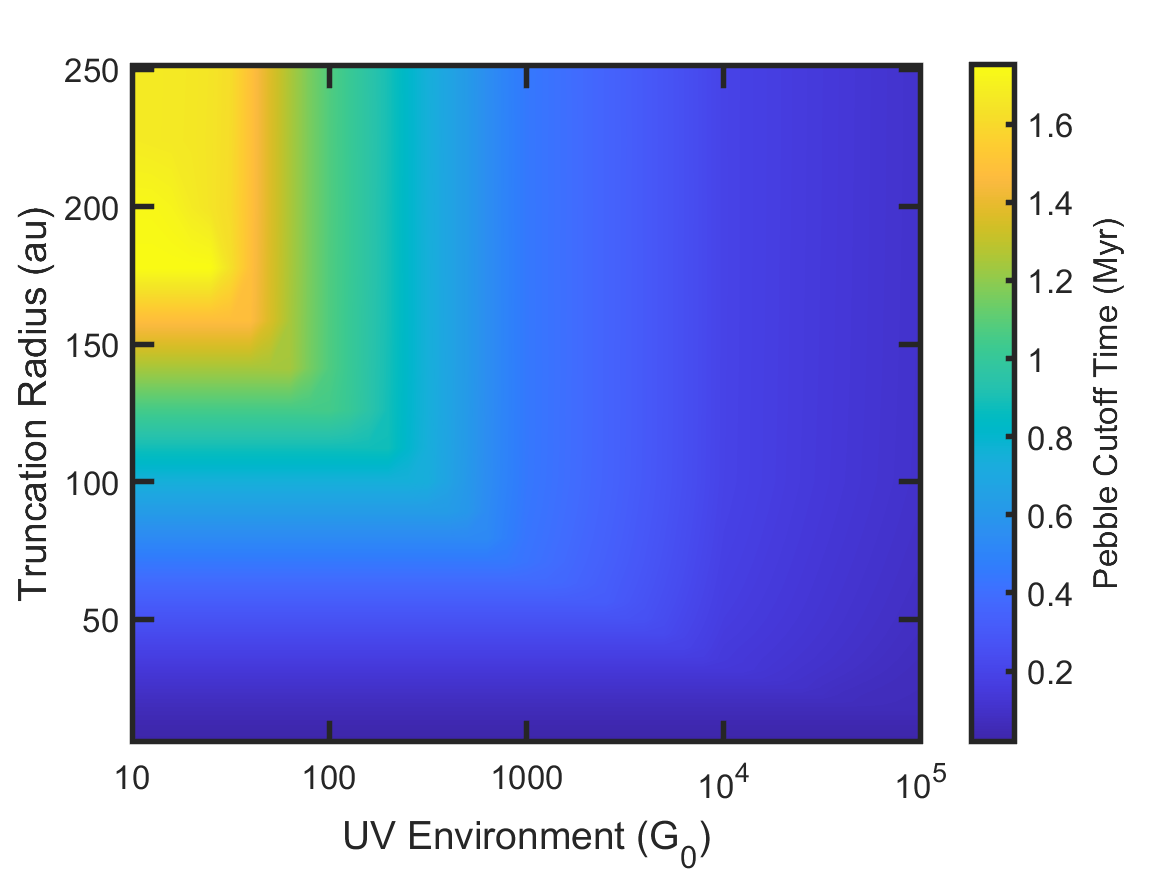}
\caption{Contour plots showing the pebble cutoff time in protoplanetary discs as a function of UV environment strength, and the truncation radius due to the outer companion. The strength of the viscous turbulent parameter was set to $\alpha=10^{-3.5}$.}
\label{fig:peb_cutoff_grid}
\end{figure}

With Figs. \ref{fig:peb_cutoff_g0} and \ref{fig:peb_cutoff_trunc} only showing the pebble cutoff times for discreet values of either the external UV radiation field or the binary truncation radius, Fig. \ref{fig:peb_cutoff_grid} shows a contour plot of the pebble cutoff time for more continuous distributions of the UV field and binary truncation radii. The strength of the viscous $\alpha$ parameter was set to $10^{-3.5}$ since there was little difference in the pebble cutoff times as a function of $\alpha$. From Fig. \ref{fig:peb_cutoff_grid} it is clear that the top left corner of the plot showing discs evolving in weak UV environments and with large binary truncation radii, have the longest pebble cutoff times, as expected from what was found in Figs. \ref{fig:peb_cutoff_g0} and \ref{fig:peb_cutoff_trunc}. As the truncation radius decreases, and the external UV field increases, the pebble cutoff times decrease, to where they are $\leq0.2$ Myr for the more extreme cases. Additionally, it is clear to see where either the external UV environment or the binary truncation radius dominates the determination of the pebble cutoff time, by either looking horizontally or vertically respectively for where the values reach convergence. The faint diagonal line, emanating from the corner of the yellow region locates where both the external UV environment and the binary truncation radius are jointly setting the pebble cutoff time, as the equilibria between viscous expansion and external photoevaporation is located at around the binary truncation radius for those combinations of $\alpha$ and the UV field. Ultimately, what is clear from Fig. \ref{fig:peb_cutoff_grid} is that the pebble cutoff time is severely reduced for lesser separated binaries, as well in stronger UV environments, with it being reduced by at least a half for discs evolving in either $\rm F_{\rm UV}\geq400 G_0$ or $r_{\rm t}\leq 100\au$.

\section{Competition between external photoevaporation and wide binary truncation}
\label{sec:comp}

With the section above showing that the pebble cutoff time is substantially affected by external photoevaporation and the truncation radius due to an external binary companion, we now determine the effects that this has on the formation potential of planetesimals and then planets. Since planetesimals are expected to form through the gravitational collapse of pebble clumps \citep{Simon16,Abod19}, the longevity of pebbles growing and drifting through the disc is therefore imperative for planetesimal formation rates. Consequently, as the initial planetary embryos are taken as the largest planetesimal that forms when a pebble cloud undergoes gravitational collapse, and additionally that the main initial growth stages for planetary embryos is through pebble and planetesimal accretion, the role of the pebble cutoff time is even more important for forming and growing planets, than it is for forming planetesimals. In this section we determine these effects on the formation of planets and planetesimals, and establish the role of the external UV environment and that of a wide binary companion.

\begin{figure}
\centering
\includegraphics[scale=0.5]{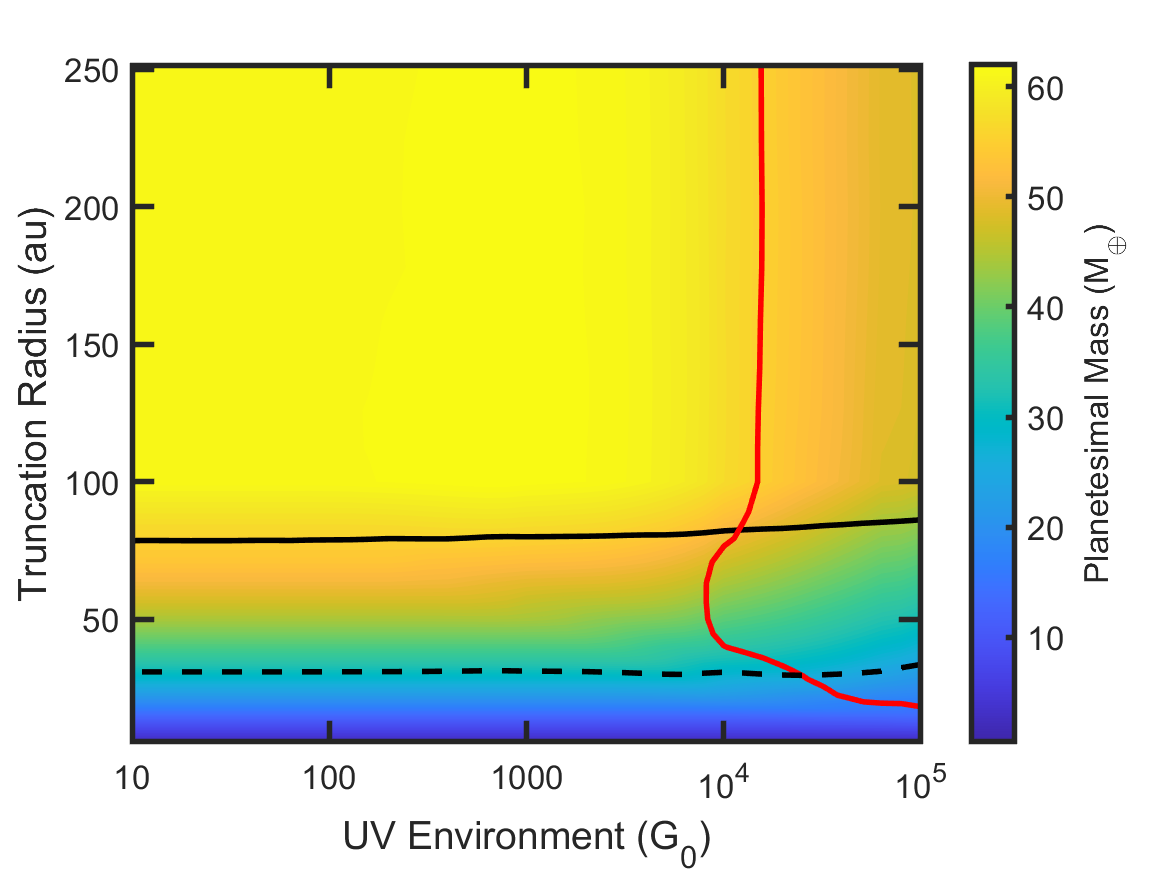}
\caption{Contour plots showing the total planetesimal mass forming in protoplanetary discs as a function of UV environment strength, and the truncation radius due to the outer companion. The strength of the viscous turbulent parameter was set to $\alpha=10^{-3.5}$. The red line shows where the total planetesimal mass was reduced to 90$\%$ of that for a disc evolving in a 10$\rm G_0$ environment, whilst the black lines show where the total planetesimal mass was reduced to 90$\%$ (solid) and $50\%$ (dashed) of that for a disc with a truncation radius of 250$\au$.}
\label{fig:pltml_masses}
\end{figure}

\subsection{Overall Planetesimal Masses}
\label{sec:pltml}
Using the planetesimal formation model outlined in \citetalias{Coleman21}, we determine the total mass of planetesimals that are able to form in our simulations with varying external UV environments and binary truncation radii. Figure \ref{fig:pltml_masses} shows the total mass of planetesimals formed in each disc as  a function of the UV field and the truncation radius. The black lines show where the planetesimal mass has been reduced to 90$\%$ (solid) and 50$\%$ (dashed) of that which formed for a disc with a truncation radius of 250$\au$, i.e. highlighting the effect of the disc truncation radius for discs in identical UV environments. Likewise, the red solid line shows where the planetesimal mass is decreased to 90$\%$ of that for a disc evolving in a 10$\rm G_0$ environment, highlighting the effect of the environment for discs evolving with identical truncation radii. Similar to Fig. \ref{fig:peb_cutoff_grid}, $\alpha$ was set to $10^{-3.5}$. Looking at the top left region of Fig. \ref{fig:pltml_masses}, where the truncation radius was set to 250$\au$, and the UV field was 10$\rm G_0$, a total of $\sim60\me$ of planetesimals were able to form throughout the entire disc. The majority of these planetesimals formed over the first 0.5 Myr of the discs lifetime, similar to that found in \citetalias{Coleman21}, and also consistent with 3D MHD simulations \citep{Huhn25}. Note due to the inside-out nature of planetesimal formation, planetesimals and planetary embryos at small orbital distances, i.e. $\sim$few$\au$, are able to form in the first 0.1 Myr of the simulations. The top left region here represents where both effects of truncation and the UV environment were at their weakest, i.e. the discs were effectively evolving in isolation.

As the truncation radius decreased or the UV field increased, the total mass of planetesimals that were able to form decreased, since the pebble production front ceased at an earlier time, resulting in fewer solids and less time for planetesimals to be able to form. However, whilst the decreases are apparent, especially in the bottom right corner of Fig. \ref{fig:pltml_masses} where the smallest truncation radius and largest UV field strength resulted in $\leq5\me$ of planetesimals being able to form, the effect on planetesimals was not severe for most of the parameter space studied. Indeed, the solid lines show where the truncation radius (black) or the UV field (red) reduce the planetesimal mass down to 90$\%$ of the maximum value that was obtained in the isolated example. This reduction occurred at truncation radii of $\sim 80\au$, corresponding to binary separations of $\sim 250\au$ for equal mass stars, and at UV fields $>10^4\rm G_0$. Going further, the dashed line shows where the planetesimal was reduced to 50$\%$, a much more severe reduction, and this only occurred for truncation radii $r_{\rm t}\sim 30\au$, corresponding to binary separations of 100$\au$ for equal mass binaries. Note that the 50$\%$ reduction due to the environment here did not occur in our parameter space and so would need stronger UV environments to manifest itself.

Therefore, Fig. \ref{fig:pltml_masses} shows that in terms of forming planetesimals, the truncation radius caused by the outer companion only becomes important when it is less than $\sim80\au$, irrespective of the UV field, whilst the environment becomes important at $>10^4\rm G_0$. The role of the environment however, is not independent of the binary separation and therefore truncation radius, since the truncation radius is able to determine the total planetesimal mass at smaller truncation radii, even in extremely strong UV environments, i.e. $10^5\rm G_0$. This is due to the external radiation being ineffective at launching winds for extremely compact discs that are the result of the extreme truncation due to the binary companion. Similar to the pebble cutoff time, when changing the value of $\alpha$ from between $10^{-4}$--$10^{-3}$, we find very few differences in terms of where the truncation radius and the UV field begin to affect the total mass of planetesimals that form. 

\begin{figure*}
\centering
\includegraphics[scale=0.4]{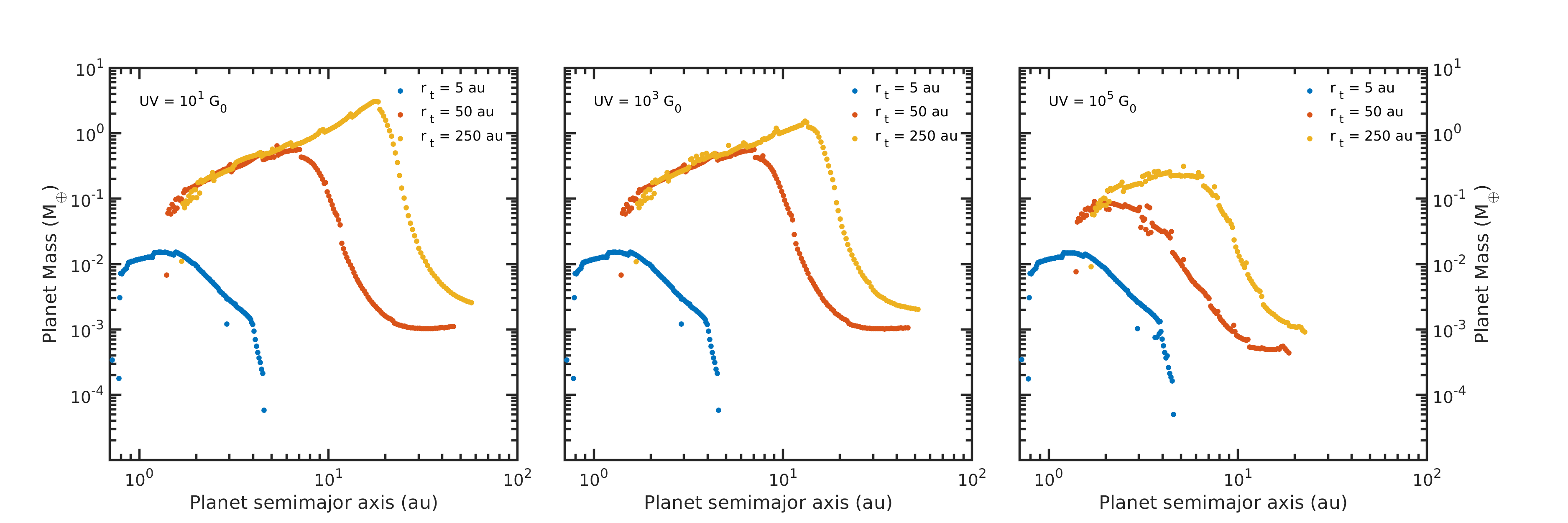}
\caption{Planet masses versus planet semimajor axis locations for specific discs. Each set of colours shows planets formed in discs with different truncation radii of $r_t = 5\au$ (blue), $r_t = 50\au$ (blue) and $r_t = 250\au$ (yellow). The different panels show the results for different UV environments of $10 \rm G_0$ (left-hand panel), $10^3 \rm G_0$ (middle panel) and $10^5 \rm G_0$ (right-hand panel). The strength of the viscous turbulent parameter was set to $\alpha=10^{-3.5}$.}
\label{fig:systems}
\end{figure*}

\subsection{Planet Growth}
\label{sec:planet}

Whilst it was shown above and in Fig. \ref{fig:pltml_masses} that the truncation radius and UV field only had moderate effects on the mass of planetesimals that were able to form, this may not be the case for the mass in planets. Following \citetalias{Coleman21}, we assume that the largest planetesimal that forms as a pebble cloud collapses is a planetary embryo. After forming, these planetary embryos then accrete pebbles and planetesimals to grow into more massive planets. Similar to \citetalias{Coleman21} we do not include any N-body effects or migration in these models, as we are mainly focussed on determining how the truncation radius and external photoevaporation affect the the amount of mass that can be converted from dust and pebbles into the planets themselves. 

Generally as the planetary embryos form, they have masses between $10^{-6}$--$10^{-3}\me$, similar to that found in \citetalias{Coleman21}. Once formed, the planetary embryos accreted pebbles until the pebble production front reached the disc outer edge, i.e. the pebble cutoff time. They were also able to accrete planetesimals, albeit at much reduced rates compared to pebble accretion, for the entirety of the disc lifetime. Figure \ref{fig:systems} shows the final planet masses versus semimajor axes that formed in different discs. The varying colours represent discs with different truncation radii of, $r_{\rm t}=5\au$ (blue), $r_{\rm t}=50\au$ (red), and $r_{\rm t}=250\au$ (yellow). The left-hand panel shows the planets that formed in discs in a $10 \rm G_0$ environment, the middle panel for a $10^3 \rm G_0$ environment, and the right-hand panel for a $10^5 \rm G_0$ environment. As can be seen in all of the panels, the number of planets, their radial extent, and their masses increase as the truncation radius increases. This is not unexpected, since the discs will be more massive and more extended, and as noted above, important time-scales such as the pebble cutoff time, occurs at later times for the discs with larger truncation radii. As the strength of the UV environment increases, in all discs apart from those with $r_{\rm t}=5\au$, the mass of the planets decreases, since external photoevaporation removes material from the disc, and more importantly, truncates the discs more than they already are due to the outer binary companions. Looking at the range in planet masses, it is also noticeable that for the more compact discs, e.g. those with $r_{\rm t}=5\au$, the maximum mass that planets reach is roughly a Lunar mass ($10^{-2}\me$). This easily highlights the difficulty in forming planetary systems in such compact binary configurations. Even for the more extended discs of $r_{\rm t}=50\au$, all planets remain under $1\me$, which falls to $0.1\me$ in high UV environments as shown in the right-hand panel. We again note that the dynamical effects of the outer binary star are not included on the growth of planets. Since most planets formed far from the outer edge of the disc, we expect these effects to be minimal. Indeed, for the majority of discs, all planets formed at distances less than half the distance to the disc outer edge, and thus the binary effects would be expected to minimal \citep{Quintana07}. However perturbations from the outer binary star may be able to affect he outer planets that form, and induce collisions between neighbouring planetary embryos. These effects will be included and explored in future work, as well as the long term evolution of formed planets.

\begin{figure}
\centering
\includegraphics[scale=0.5]{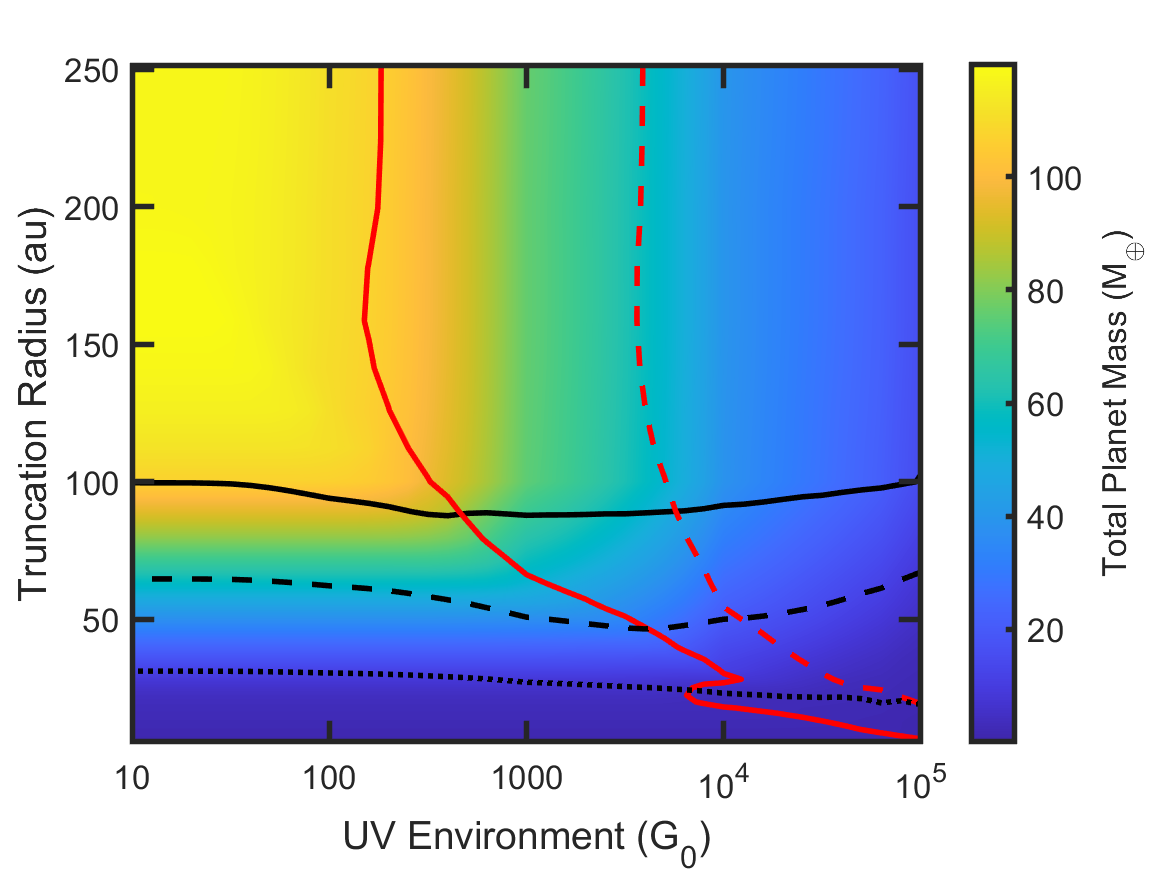}
\caption{Contour plots showing the total planets mass forming in protoplanetary discs as a function of UV environment strength, and the truncation radius due to the outer companion. The strength of the viscous turbulent parameter was set to $\alpha=10^{-3.5}$. The red lines show where the total planetesimal mass was reduced to 90$\%$ (solid) and 50$\%$ (dashed) of that for a disc evolving in a 10$\rm G_0$ environment, whilst the black lines show where the total planetesimal mass was reduced to 90$\%$ (solid), $50\%$ (dashed) and $10\%$ (dotted) of that for a disc with a truncation radius of 250$\au$.}
\label{fig:planet_mass}
\end{figure}

Whilst Fig. \ref{fig:systems} showed example systems that form in systems of different binary separations, or in different UV environments, in Fig. \ref{fig:planet_mass} we show the total planet masses for each simulation as a function of the UV field and the truncation radius. Similar to Fig. \ref{fig:pltml_masses}, the solid, dashed and dotted lines show where the total planet mass has been reduced to 90$\%$, $50\%$ and $10\%$ of the maximum values that form as a function of the UV field (red) and the truncation radius (black) respectively. When looking at where the maximum total planet mass was found, this occurred in the top left region of Fig. \ref{fig:planet_mass} where akin to the maximum planetesimal mass above, this was for the most extended discs and in the weakest UV environments. For the most isolated disc, the combined mass of all the planets that were able to form and grow there totalled $\sim115\me$, with individual planets reaching masses up to $m_{\rm p}=3\me$.
The distribution of these planet's masses as a function of semimajor axis can be seen by the yellow points in the left-hand panel of Fig. \ref{fig:systems}. Should migration and N-body effects be accounted for, including those due to the outer binary star, then it would be expected that these planets would have been able to interact and collide, growing into more massive planets.

Looking at the effects of the truncation radius on the total planet mass, they are slightly more extreme than the effects on the total planetesimal mass. Whilst for planetesimals, their total masses began to be depleted for truncation radii less than 80$\au$, for planets, this transition occurred at 100 $\au$, irrespective of the UV field. This is due to pebble accretion being the main mechanism for planetary embryos to gain mass and grow from their formation masses of $m_{\rm p}\le 10^{-3}\me$ up to an Earth mass and beyond. Since this takes some time, it is only natural that the transition for the reduction in total planet mass moves to larger truncation radii, with the difference effectively being the amount of time required for the planetary embryos to grow into planets before the pebble production front reaches the disc outer edge. This is also the case when determining the effects of the UV field, where the reduction to 90$\%$ of the total planet mass formed in a $10\rm G_0$ environment, now occurs at $\sim 200\rm G_0$, instead of $10^4\rm G_0$ that was found for the total planetesimal mass. This shows how the truncation of the disc in even those moderate UV environments can have an effect on planetary growth, as seen in other works \citep{Winter22,Qiao23,Qiao26}.

\begin{figure}
\centering
\includegraphics[scale=0.5]{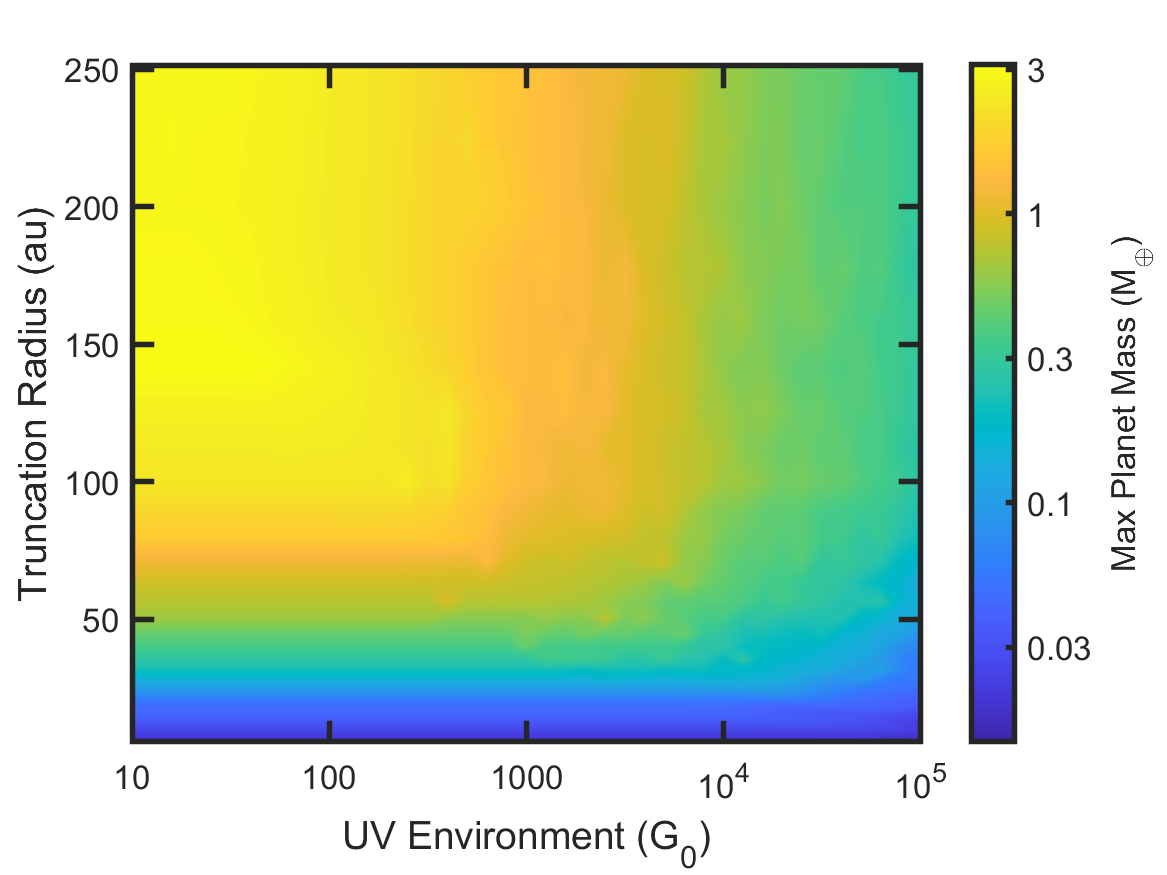}
\caption{Contour plots showing the maximum mass of planets forming in protoplanetary discs as a function of UV environment strength, and the truncation radius due to the outer companion. The strength of the viscous turbulent parameter was set to $\alpha=10^{-3.5}$.}
\label{fig:max_planet_mass}
\end{figure}

Moving to more extreme reductions down to $50\%$ of the maximum total mass formed, as shown by the dashed lines in Fig. \ref{fig:planet_mass}, we can see that these occur at $r_{\rm t}\sim50$--$70\au$, and for UV fields of $\sim 4000 \rm G_0$. This corresponds to binary separations of between 150--200$\au$ for equal mass binaries. For the most severe reductions, showed by the dotted line, where the reduction is down to $10\%$, this is seen to occur at truncation radii of 20--30$\au$, corresponding to binary separations of between 60--90$\au$ for equal mass binaries. This indicates that for more compact binaries, planet formation is severely hindered by the outer companion, whilst for larger binary separations, the effects are minimal. This is consistent with observations where for binary separations between 100--300$\au$, the planetary systems are mostly similar to those found in single star systems, whilst for separations $<100\au$, there is a dearth of multi planet systems as well as fewer individual giant planet systems \citep{Roell12,Su21,Hirsch21}. Indeed, \citet{Hirsch21} find that 20\% of binaries with separations larger than 100 $\au$ contain giant planets, similar to that found around single stars, whilst for separations less than 100 $\au$, this drops to $\sim4\%$. The results here show that with planet formation hindered for binaries with separations less than 100 $\au$, it would be difficult to form the giant planet cores required to undergo runaway gas accretion and become giant planets, providing a possible explanation for the reduction in observed giant planet occurrence rates as a function of the binary separation.

Interestingly, when looking at the maximum planet mass that is able to form in the simulations, though without any N-body or migration effects, Fig. \ref{fig:max_planet_mass} shows that for truncation radii of $r_{\rm t}\le 75\au$, no planets above an Earth mass are able to form, which could reduce the effectiveness of collisions in the outer disc, since those planets would be more dynamically separated, further indicating for those corresponding binary separations, planet formation is significantly reduced. This is also seen to occur for UV fields $\ge 2000\rm G_0$, again showing where external photoevaporation begins to dictate planet formation processes.

\begin{figure}
\centering
\includegraphics[scale=0.5]{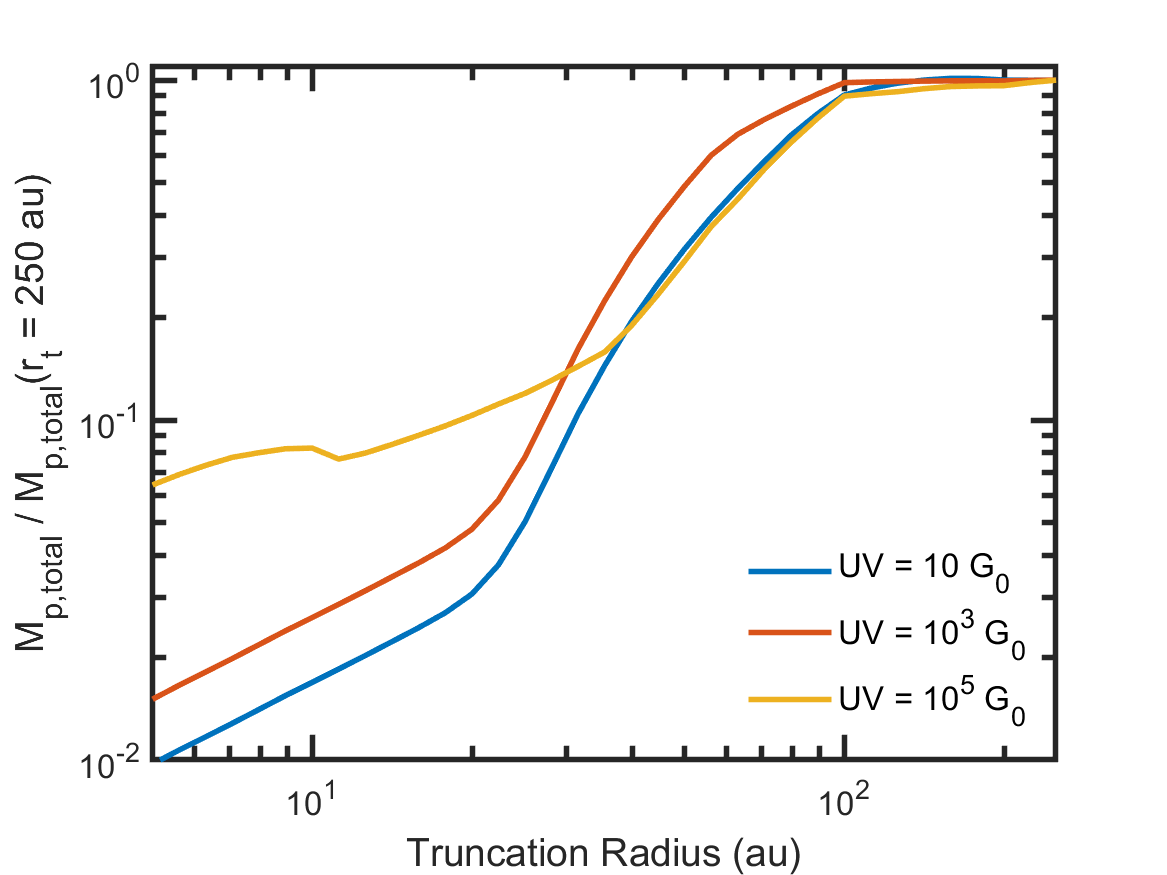}
\caption{Ratio of total planet mass formed in discs with different binary truncation radii compared to the total planet mass formed in a disc with a truncation radii of 250 $\au$. The UV field strength is denoted by the line colours where we show for $10\rm G_0$ (blue), $10^3\rm G_0$ (red) and $10^5\rm G_0$ (yellow).}
\label{fig:total_rt}
\end{figure}

\begin{figure}
\centering
\includegraphics[scale=0.5]{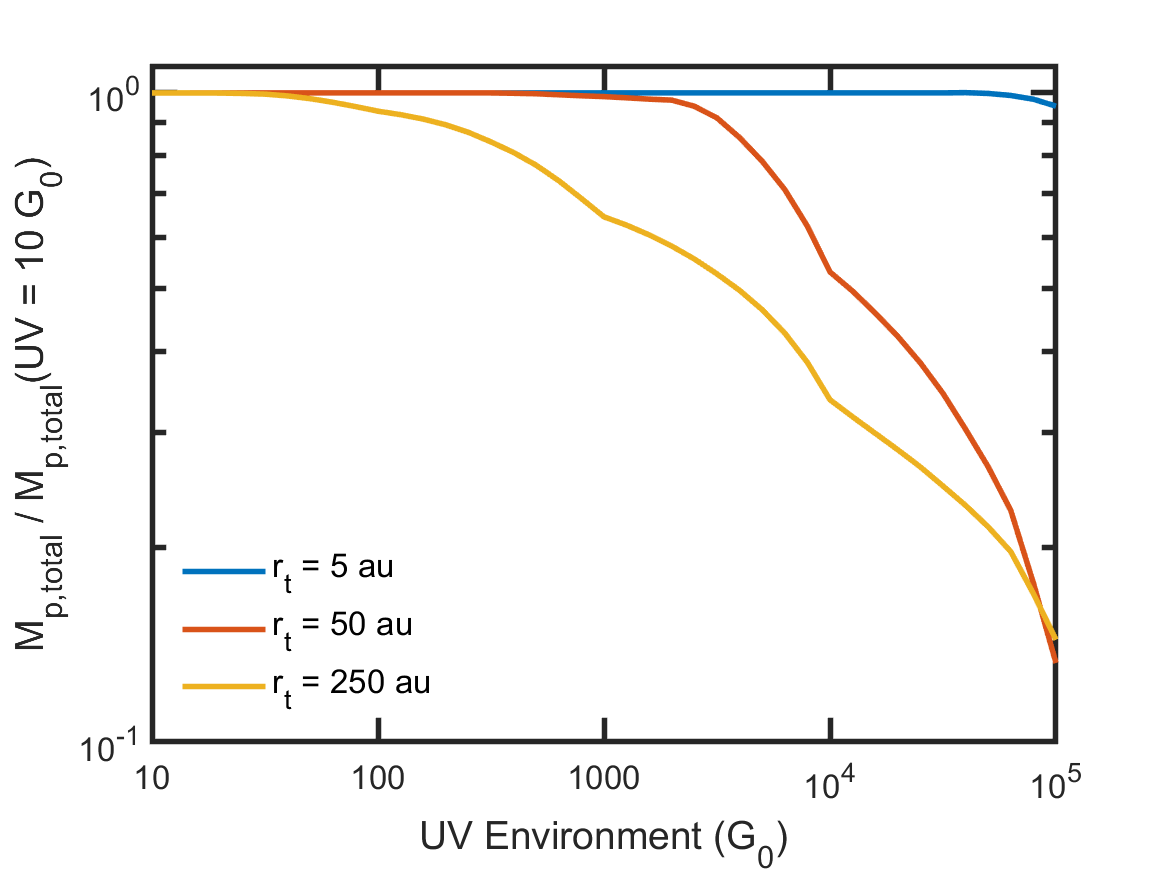}
\caption{Ratio of total planet mass formed in discs evolving in different UV environments compared to the total planet mass formed in a disc evolving in a UV environment of $10\rm G_0$. The truncation radii is denoted by the line colours where we show for $r_t = 5\au$ (blue), $r_t = 50\au$ (red) and $r_t = 250\au$ (yellow).}
\label{fig:total_uv}
\end{figure}

Additionally, similar to what was found in Fig. \ref{fig:pltml_masses}, the effect of the truncation radius appear roughly independent of the UV field for the parameters tested here. This is easily seen in Fig. \ref{fig:total_rt} which shows the ratio of the total mass formed in each simulation compared to that formed in a disc with a truncation radius of $250\au$. The colours highlight discs evolving in different UV environments, where it is clear that apart from for small truncation radii, i.e. $<30\au$, the ratios are similar for UV fields between 10--$10^5\rm G_0$. It is also clear that for all of the environments studied here, planet formation is reduced to the $10\%$ level for truncation radii $r_{\rm t}\sim20$--$30\au$.

Whilst the ratio of total masses as a function of the truncation radii appeared roughly independent of the UV field, it is not quite the same in the opposite case. Figure \ref{fig:total_uv} shows the ratio of total planet mass formed in discs in different UV environments, compared to that for a disc in a $10\rm G_0$ environment, with the colours representing different truncation radii. For large discs, where the truncation radius is large, e.g. 250$\au$, then the ratio steadily decreases, reaching $\sim50\%$ at $\sim 4000\rm G_0$, before approaching $10\%$ at $10^5 \rm G_0$. For small discs, e.g. $r_{\rm t} = 5\au$, as shown by the blue line, the ratio remains nearly constant, since the discs were already sufficiently truncated and so the external radiation was not able to launch effective photoevaporative winds. This shows that for small truncation radii, or binary separations, it is those that will dominate the evolution of planets, whereas for the wider separation binaries, external photoevaporation plays a key, and increasingly important role.

\section{Discussion and Conclusions}
\label{sec:conc}

In this paper, we have explored how the formation of planetesimals and planetary embryos is affected by an outer companion in a wide binary system, as well as by the effects of an external UV radiation field due to nearby massive stars. We ran numerous protoplanetary disc evolution models, spanning a range of truncation radii that represents larger variations in binary separations and eccentricities, a range in the viscous turbulent parameter $\alpha$, and a range of UV environments that determine external photoevaporative mass-loss rates. We have determined for Solar mass stars, where the effects of an outer binary companion, and at what strength of the local UV environments, that planet formation rates are significantly reduced. This is due to planetary embryos not having enough time to accrete significant amounts of solids so that they can evolve in to planets. We draw the following main conclusions from this work.

(1) We find that both small truncation radii and strong UV environments result in shorter pebble cutoff times, the time at which the pebble production front reaches the disc outer edge. These reductions begin to occur for truncation radii less than 200$\au$, and in UV environments greater than $30 \rm G_0$, where they start reducing from their nominal values of $\sim2$ Myr down to $\sim0.1$ Myr for the strongest UV environments and smallest truncation radii.

(2) The shorter pebble cutoff times resulted in reduced overall planetesimal masses being able to form in discs with smaller truncation radii. Compared to discs with large binary separations and in weak UV environments, i.e. the maximum case, overall planetesimal masses were reduced to $90\%$ for truncation radii less than 80$\au$, and down to $50\%$ for truncation radii less than 30$\au$. These correspond to binary separations of $\sim 250\au$ and $\sim 90\au$ respectively.

(3) The strength of the UV environment also reduces the overall planetesimal masses, but to a lesser extent than the truncation radii. For discs evolving in UV environments greater than $10^4\rm G_0$, overall planetesimal masses are reduced to 90$\%$ of that for discs evolving in weak UV environments, e.g. $10\rm G_0$.

(4) The effects on planet masses were more pronounced than that on planetesimal masses, due to the reduction in the time available that planets could accrete pebbles. For truncation radii less than $\sim30\au$, planet masses were reduced to $10\%$ of that for forming around single stars. For equal mass binaries, planet formation is severely reduced for separations less than $\sim100\au$, consistent with observed reductions in giant planets and multiple planetary systems among exoplanet observations \citep{Roell12,Su21}.

(5) The UV environment strength also reduces the total planet mass, with UV fields of $\sim200\rm G_0$ and $\sim 4000\rm G_0$ reducing total planet masses down to $90\%$ and $50\%$ of that which can form in weak UV environments, when the effects of external photoevaporation are negligible.

(6) Finally, the effects of the binary truncation radius are more prominent than that from the UV environment, except for in extremely strong UV environments, or when the binary separation is large, and thus having limited effects on the truncation of the discs.

The simulations here show that the truncation of protoplanetary discs by either the external UV environment, and/or the perturbative effects of an outer binary companion, significantly reduce the planet forming potential of circumstellar discs.
The effects of the outer companion are especially extreme for separations $<100\au$ for equal mass binaries, since significantly fewer planetesimals and planetary embryos are able to form and grow. This reduces the possibilities for giant planet cores to be able to form, which can be an explanation for the reduction in giant planet occurrence rates as binary separations decreases \citep{Hirsch21}. We also found that the environment is important for moderate UV fields ($\sim10^3\rm G_0$). Additionally this work also highlights the importance of including the formation of planetesimals and planetary embryos in planet formation models, especially when incorporating the effects of the local environment, be that the radiation of nearby massive stars, or the truncation effects of a binary companion.

There are also a number of other effects that can affect the speed at which the pebble production front moves through the disc, as well as how effective external photoevaporation is at truncating protoplanetary discs. In this work, we only considered Solar metallicity, however it has been shown that higher metallicities can allow smaller discs to form more massive planets from initially formed planet seeds \citep{Savvidou23}. It is unclear whether this would be the case here, since the increased metallicity would result in the pebble production front evolving faster, thus reducing the pebble cutoff time. With that occurring, the planetary embryos would have less time to accrete pebbles in order to reach efficient growth regimes where they can form into more massive planets, rendering the effects of metallicity unclear. Whilst this was not explored here, this will be explored in future work, including the effects of other planet formation processes such as gas accretion and planet migration. With the pebble cutoff time depending on the external photoevaporation rate in stronger UV environments, there are a number of things that can act to reduce their effectiveness. These include the microphysical properties such as whether the dust in the wind has undergone grain growth \citep{Coleman25}, as well as shielding from the surrounding disc \citep{Qiao22,Qiao23,Qiao26}. The effects of shielding can be extremely effective, as they allow discs that would otherwise be in strong UV environments, to evolve for the early portion of their lifetimes in weaker environments, aiding disc longevity as well the growth of planets \citep{Qiao22,Qiao23}. In regards to this work, should shielding be included, this would diminish the effects on disc truncation due to external photoevaporation to an extent, allowing the effects due to the binary star to dominate for a larger region of the parameter space. This would obviously depend on the length of the shielding time, and should be explored in future work.

Whilst this work demonstrated the effects of a wide binary companion on the formation potential of planets in wide binary systems through pebble and planetesimal accretion, it did not include other physical processes that can further affect the evolution of the planets and the final architecture of the planetary systems. These processes include N-body interactions between the growing planets as well as secular perturbations from the binary companion, disc--planet interactions that result in planetary migration \citep{pdk10,pdk11,LinPapaloizou86}, as well as accretion of gas on to embedded planets \citep{CPN17,Poon21}. These processes would be expected to allow planets to increase in mass, since they may be able to migrate into regions where planetesimal populations are plentiful, accrete significant abundances of gas, and grow through mutual collisions with other planetary embryos. In future work, we will include these processes and create a global model of planet formation for wide binary systems. This model will then be used to explore the formation and evolution of planets in wide binary systems as well as those planets that will undoubtedly be ejected from their systems through increased gravitational interactions in the system due to the outer companions, becoming free floating planets. The development of such a model will compliment pre-existing global models of planet formation around single stars, and within circumbinary systems, and together will provide planet formation scenarios for all planets forming across the galaxy. The comparisons of the results of such models with observed systems and the distributions of known planets will determine where future efforts need to lie in order to better understand the planet formation framework.

\section*{Data Availability}
The data underlying this article will be shared on reasonable request to the corresponding author.

\section*{Acknowledgements}
We thank the anonymous referee for their useful and insightful comments that improved the quality of the paper.
GALC acknowledges funding from the UKRI/STFC grant ST/X000931/1.
This research utilised Queen Mary's Apocrita HPC facility, supported by QMUL Research-IT (http://doi.org/10.5281/zenodo.438045).

\bibliographystyle{mnras}
\bibliography{references}{}

\begin{thebibliography}{}
\makeatletter
\relax
\def\mn@urlcharsother{\let\do\@makeother \do\$\do\&\do\#\do\^\do\_\do\%\do\~}
\def\mn@doi{\begingroup\mn@urlcharsother \@ifnextchar [ {\mn@doi@} {\mn@doi@[]}}
\def\mn@doi@[#1]#2{\def\@tempa{#1}\ifx\@tempa\@empty \href {http://dx.doi.org/#2} {doi:#2}\else \href {http://dx.doi.org/#2} {#1}\fi \endgroup}
\def\mn@eprint#1#2{\mn@eprint@#1:#2::\@nil}
\def\mn@eprint@arXiv#1{\href {http://arxiv.org/abs/#1} {{\tt arXiv:#1}}}
\def\mn@eprint@dblp#1{\href {http://dblp.uni-trier.de/rec/bibtex/#1.xml} {dblp:#1}}
\def\mn@eprint@#1:#2:#3:#4\@nil{\def\@tempa {#1}\def\@tempb {#2}\def\@tempc {#3}\ifx \@tempc \@empty \let \@tempc \@tempb \let \@tempb \@tempa \fi \ifx \@tempb \@empty \def\@tempb {arXiv}\fi \@ifundefined {mn@eprint@\@tempb}{\@tempb:\@tempc}{\expandafter \expandafter \csname mn@eprint@\@tempb\endcsname \expandafter{\@tempc}}}

\bibitem[\protect\citeauthoryear{{Abod}, {Simon}, {Li}, {Armitage}, {Youdin}  \& {Kretke}}{{Abod} et~al.}{2019}]{Abod19}
{Abod} C.~P.,  {Simon} J.~B.,  {Li} R.,  {Armitage} P.~J.,  {Youdin} A.~N.,   {Kretke} K.~A.,  2019, \mn@doi [\apj] {10.3847/1538-4357/ab40a3}, \href {https://ui.adsabs.harvard.edu/abs/2019ApJ...883..192A} {883, 192}

\bibitem[\protect\citeauthoryear{{Adachi}, {Hayashi}  \& {Nakazawa}}{{Adachi} et~al.}{1976}]{Adachi}
{Adachi} I.,  {Hayashi} C.,   {Nakazawa} K.,  1976, \mn@doi [Progress of Theoretical Physics] {10.1143/PTP.56.1756}, \href {http://adsabs.harvard.edu/abs/1976PThPh..56.1756A} {56, 1756}

\bibitem[\protect\citeauthoryear{{Adams}, {Hollenbach}, {Laughlin}  \& {Gorti}}{{Adams} et~al.}{2004}]{Adams04}
{Adams} F.~C.,  {Hollenbach} D.,  {Laughlin} G.,   {Gorti} U.,  2004, \mn@doi [\apj] {10.1086/421989}, \href {https://ui.adsabs.harvard.edu/abs/2004ApJ...611..360A} {611, 360}

\bibitem[\protect\citeauthoryear{{Alibert} et~al.,}{{Alibert} et~al.}{2006}]{Alibert2006}
{Alibert} Y.,  et~al., 2006, \mn@doi [\aap] {10.1051/0004-6361:20065697}, \href {http://adsabs.harvard.edu/abs/2006A%26A...455L..25A} {455, L25}

\bibitem[\protect\citeauthoryear{{Allen} et~al.,}{{Allen} et~al.}{2025}]{PastPresentFuture2025}
{Allen} M.,  et~al., 2025, \mn@doi [The Open Journal of Astrophysics] {10.33232/001c.137538}, \href {https://ui.adsabs.harvard.edu/abs/2025OJAp....8E..54A} {8, 54}

\bibitem[\protect\citeauthoryear{{Anania} et~al.,}{{Anania} et~al.}{2025}]{Anania25}
{Anania} R.,  et~al., 2025, \mn@doi [\apj] {10.3847/1538-4357/adb587}, \href {https://ui.adsabs.harvard.edu/abs/2025ApJ...989....8A} {989, 8}

\bibitem[\protect\citeauthoryear{{Anglada-Escud{\'e}} et~al.,}{{Anglada-Escud{\'e}} et~al.}{2016}]{Anglada2016}
{Anglada-Escud{\'e}} G.,  et~al., 2016, \mn@doi [\nat] {10.1038/nature19106}, \href {http://adsabs.harvard.edu/abs/2016Natur.536..437A} {536, 437}

\bibitem[\protect\citeauthoryear{{Ansdell}, {Williams}, {Manara}, {Miotello}, {Facchini}, {van der Marel}, {Testi}  \& {van Dishoeck}}{{Ansdell} et~al.}{2017}]{Ansdell17}
{Ansdell} M.,  {Williams} J.~P.,  {Manara} C.~F.,  {Miotello} A.,  {Facchini} S.,  {van der Marel} N.,  {Testi} L.,   {van Dishoeck} E.~F.,  2017, \mn@doi [\aj] {10.3847/1538-3881/aa69c0}, \href {https://ui.adsabs.harvard.edu/abs/2017AJ....153..240A} {153, 240}

\bibitem[\protect\citeauthoryear{{Ansdell} et~al.,}{{Ansdell} et~al.}{2018}]{Ansdell18}
{Ansdell} M.,  et~al., 2018, \mn@doi [\apj] {10.3847/1538-4357/aab890}, \href {https://ui.adsabs.harvard.edu/abs/2018ApJ...859...21A} {859, 21}

\bibitem[\protect\citeauthoryear{{Arlt} \& {Urpin}}{{Arlt} \& {Urpin}}{2004}]{Arlt04}
{Arlt} R.,  {Urpin} V.,  2004, \mn@doi [\aap] {10.1051/0004-6361:20035896}, \href {https://ui.adsabs.harvard.edu/abs/2004A&A...426..755A} {426, 755}

\bibitem[\protect\citeauthoryear{{Artymowicz} \& {Lubow}}{{Artymowicz} \& {Lubow}}{1994}]{Artymowicz94}
{Artymowicz} P.,  {Lubow} S.~H.,  1994, \mn@doi [\apj] {10.1086/173679}, \href {https://ui.adsabs.harvard.edu/abs/1994ApJ...421..651A} {421, 651}

\bibitem[\protect\citeauthoryear{{Ataiee}, {Baruteau}, {Alibert}  \& {Benz}}{{Ataiee} et~al.}{2018}]{Ataiee18}
{Ataiee} S.,  {Baruteau} C.,  {Alibert} Y.,   {Benz} W.,  2018, \mn@doi [\aap] {10.1051/0004-6361/201732026}, \href {https://ui.adsabs.harvard.edu/abs/2018A&A...615A.110A} {615, A110}

\bibitem[\protect\citeauthoryear{{Bai} \& {Stone}}{{Bai} \& {Stone}}{2014}]{Bai2014}
{Bai} X.-N.,  {Stone} J.~M.,  2014, \mn@doi [\apj] {10.1088/0004-637X/796/1/31}, \href {http://adsabs.harvard.edu/abs/2014ApJ...796...31B} {796, 31}

\bibitem[\protect\citeauthoryear{{Barbieri}, {Marzari}  \& {Scholl}}{{Barbieri} et~al.}{2002}]{Barbieri02}
{Barbieri} M.,  {Marzari} F.,   {Scholl} H.,  2002, \mn@doi [\aap] {10.1051/0004-6361:20021357}, \href {https://ui.adsabs.harvard.edu/abs/2002A&A...396..219B} {396, 219}

\bibitem[\protect\citeauthoryear{{B{\'e}thune}, {Lesur}  \& {Ferreira}}{{B{\'e}thune} et~al.}{2016}]{BethuneLesur2016}
{B{\'e}thune} W.,  {Lesur} G.,   {Ferreira} J.,  2016, \mn@doi [\aap] {10.1051/0004-6361/201527874}, \href {http://adsabs.harvard.edu/abs/2016A%26A...589A..87B} {589, A87}

\bibitem[\protect\citeauthoryear{{Bitsch}, {Lambrechts}  \& {Johansen}}{{Bitsch} et~al.}{2015}]{Bitsch15}
{Bitsch} B.,  {Lambrechts} M.,   {Johansen} A.,  2015, \mn@doi [\aap] {10.1051/0004-6361/201526463}, \href {http://adsabs.harvard.edu/abs/2015A%26A...582A.112B} {582, A112}

\bibitem[\protect\citeauthoryear{{Bitsch}, {Morbidelli}, {Johansen}, {Lega}, {Lambrechts}  \& {Crida}}{{Bitsch} et~al.}{2018}]{Bitsch18}
{Bitsch} B.,  {Morbidelli} A.,  {Johansen} A.,  {Lega} E.,  {Lambrechts} M.,   {Crida} A.,  2018, \mn@doi [\aap] {10.1051/0004-6361/201731931}, \href {https://ui.adsabs.harvard.edu/abs/2018A&A...612A..30B} {612, A30}

\bibitem[\protect\citeauthoryear{{Br{\"u}gger}, {Burn}, {Coleman}, {Alibert}  \& {Benz}}{{Br{\"u}gger} et~al.}{2020}]{Brugger20}
{Br{\"u}gger} N.,  {Burn} R.,  {Coleman} G.~A.~L.,  {Alibert} Y.,   {Benz} W.,  2020, \mn@doi [\aap] {10.1051/0004-6361/202038042}, \href {https://ui.adsabs.harvard.edu/abs/2020A&A...640A..21B} {640, A21}

\bibitem[\protect\citeauthoryear{{Campbell}, {Walker}  \& {Yang}}{{Campbell} et~al.}{1988}]{Campbell88}
{Campbell} B.,  {Walker} G.~A.~H.,   {Yang} S.,  1988, \mn@doi [\apj] {10.1086/166608}, \href {https://ui.adsabs.harvard.edu/abs/1988ApJ...331..902C} {331, 902}

\bibitem[\protect\citeauthoryear{{Carrera}, {Simon}, {Li}, {Kretke}  \& {Klahr}}{{Carrera} et~al.}{2021}]{Carrera21}
{Carrera} D.,  {Simon} J.~B.,  {Li} R.,  {Kretke} K.~A.,   {Klahr} H.,  2021, \mn@doi [\aj] {10.3847/1538-3881/abd4d9}, \href {https://ui.adsabs.harvard.edu/abs/2021AJ....161...96C} {161, 96}

\bibitem[\protect\citeauthoryear{{Chambers}}{{Chambers}}{2006}]{Chambers06}
{Chambers} J.,  2006, \mn@doi [\icarus] {10.1016/j.icarus.2005.10.017}, \href {https://ui.adsabs.harvard.edu/abs/2006Icar..180..496C} {180, 496}

\bibitem[\protect\citeauthoryear{{Clarke}, {Gendrin}  \& {Sotomayor}}{{Clarke} et~al.}{2001}]{Clarke2001}
{Clarke} C.~J.,  {Gendrin} A.,   {Sotomayor} M.,  2001, \mn@doi [\mnras] {10.1046/j.1365-8711.2001.04891.x}, \href {http://adsabs.harvard.edu/abs/2001MNRAS.328..485C} {328, 485}

\bibitem[\protect\citeauthoryear{{Coleman}}{{Coleman}}{2021}]{Coleman21}
{Coleman} G. A.~L.,  2021, \mn@doi [\mnras] {10.1093/mnras/stab1904}, \href {https://ui.adsabs.harvard.edu/abs/2021MNRAS.506.3596C} {506, 3596}

\bibitem[\protect\citeauthoryear{{Coleman}}{{Coleman}}{2024}]{Coleman24FFP}
{Coleman} G. A.~L.,  2024, \mn@doi [\mnras] {10.1093/mnras/stae903}, \href {https://ui.adsabs.harvard.edu/abs/2024MNRAS.530..630C} {530, 630}

\bibitem[\protect\citeauthoryear{{Coleman} \& {Haworth}}{{Coleman} \& {Haworth}}{2022}]{Coleman22}
{Coleman} G. A.~L.,  {Haworth} T.~J.,  2022, \mn@doi [MNRAS] {10.1093/mnras/stac1513}, \href {https://ui.adsabs.harvard.edu/abs/2022MNRAS.514.2315C} {514, 2315}

\bibitem[\protect\citeauthoryear{{Coleman} \& {Nelson}}{{Coleman} \& {Nelson}}{2014}]{ColemanNelson14}
{Coleman} G.~A.~L.,  {Nelson} R.~P.,  2014, \mn@doi [\mnras] {10.1093/mnras/stu1715}, \href {http://adsabs.harvard.edu/abs/2014MNRAS.445..479C} {445, 479}

\bibitem[\protect\citeauthoryear{{Coleman} \& {Nelson}}{{Coleman} \& {Nelson}}{2016a}]{ColemanNelson16}
{Coleman} G.~A.~L.,  {Nelson} R.~P.,  2016a, \mn@doi [\mnras] {10.1093/mnras/stw149}, \href {http://adsabs.harvard.edu/abs/2016MNRAS.457.2480C} {457, 2480}

\bibitem[\protect\citeauthoryear{{Coleman} \& {Nelson}}{{Coleman} \& {Nelson}}{2016b}]{ColemanNelson16b}
{Coleman} G.~A.~L.,  {Nelson} R.~P.,  2016b, \mn@doi [\mnras] {10.1093/mnras/stw1177}, \href {http://adsabs.harvard.edu/abs/2016MNRAS.460.2779C} {460, 2779}

\bibitem[\protect\citeauthoryear{{Coleman}, {Nelson}, {Paardekooper}, {Dreizler}, {Giesers}  \& {Anglada-Escud{\'e}}}{{Coleman} et~al.}{2017a}]{ColemanProxima17}
{Coleman} G.~A.~L.,  {Nelson} R.~P.,  {Paardekooper} S.~J.,  {Dreizler} S.,  {Giesers} B.,   {Anglada-Escud{\'e}} G.,  2017a, \mn@doi [\mnras] {10.1093/mnras/stx169}, \href {http://adsabs.harvard.edu/abs/2017MNRAS.467..996C} {467, 996}

\bibitem[\protect\citeauthoryear{{Coleman}, {Papaloizou}  \& {Nelson}}{{Coleman} et~al.}{2017b}]{CPN17}
{Coleman} G.~A.~L.,  {Papaloizou} J.~C.~B.,   {Nelson} R.~P.,  2017b, \mn@doi [\mnras] {10.1093/mnras/stx1297}, \href {http://adsabs.harvard.edu/abs/2017MNRAS.470.3206C} {470, 3206}

\bibitem[\protect\citeauthoryear{{Coleman}, {Leleu}, {Alibert}  \& {Benz}}{{Coleman} et~al.}{2019}]{Coleman19}
{Coleman} G.~A.~L.,  {Leleu} A.,  {Alibert} Y.,   {Benz} W.,  2019, \mn@doi [\aap] {10.1051/0004-6361/201935922}, \href {https://ui.adsabs.harvard.edu/abs/2019A&A...631A...7C} {631, A7}

\bibitem[\protect\citeauthoryear{{Coleman}, {Nelson}  \& {Triaud}}{{Coleman} et~al.}{2023}]{Coleman23}
{Coleman} G. A.~L.,  {Nelson} R.~P.,   {Triaud} A. H.~M.~J.,  2023, \mn@doi [\mnras] {10.1093/mnras/stad833}, \href {https://ui.adsabs.harvard.edu/abs/2023MNRAS.522.4352C} {522, 4352}

\bibitem[\protect\citeauthoryear{{Coleman}, {Nelson}, {Triaud}  \& {Standing}}{{Coleman} et~al.}{2024a}]{Coleman24}
{Coleman} G. A.~L.,  {Nelson} R.~P.,  {Triaud} A. H.~M.~J.,   {Standing} M.~R.,  2024a, \mn@doi [\mnras] {10.1093/mnras/stad3216}, \href {https://ui.adsabs.harvard.edu/abs/2024MNRAS.527..414C} {527, 414}

\bibitem[\protect\citeauthoryear{{Coleman}, {Mroueh}  \& {Haworth}}{{Coleman} et~al.}{2024b}]{Coleman24MHD}
{Coleman} G. A.~L.,  {Mroueh} J.~K.,   {Haworth} T.~J.,  2024b, \mn@doi [\mnras] {10.1093/mnras/stad3692}, \href {https://ui.adsabs.harvard.edu/abs/2024MNRAS.527.7588C} {527, 7588}

\bibitem[\protect\citeauthoryear{{Coleman}, {Haworth}  \& {Qiao}}{{Coleman} et~al.}{2025a}]{Coleman25}
{Coleman} G. A.~L.,  {Haworth} T.~J.,   {Qiao} L.,  2025a, \mn@doi [\mnras] {10.1093/mnras/staf555}, \href {https://ui.adsabs.harvard.edu/abs/2025MNRAS.539.1190C} {539, 1190}

\bibitem[\protect\citeauthoryear{{Coleman}, {Kim}, {Haworth}, {Hartman}  \& {Kalish}}{{Coleman} et~al.}{2025b}]{Coleman25FUV}
{Coleman} G. A.~L.,  {Kim} J.~S.,  {Haworth} T.~J.,  {Hartman} P.~A.,   {Kalish} T.~C.,  2025b, \mn@doi [\mnras] {10.1093/mnras/staf789}, \href {https://ui.adsabs.harvard.edu/abs/2025MNRAS.540.1202C} {540, 1202}

\bibitem[\protect\citeauthoryear{{Coleman}, {Haworth}  \& {Kim}}{{Coleman} et~al.}{2025c}]{Coleman25SigOri}
{Coleman} G. A.~L.,  {Haworth} T.~J.,   {Kim} J.~S.,  2025c, \mn@doi [\mnras] {10.1093/mnrasl/slaf100}, \href {https://ui.adsabs.harvard.edu/abs/2025MNRAS.544L..70C} {544, L70}

\bibitem[\protect\citeauthoryear{{Coleman}, {Haworth}, {Schroetter}  \& {Bern{\'e}}}{{Coleman} et~al.}{2026}]{Coleman26d203}
{Coleman} G. A.~L.,  {Haworth} T.~J.,  {Schroetter} I.,   {Bern{\'e}} O.,  2026, \mn@doi [\mnras] {10.1093/mnras/staf2015}, \href {https://ui.adsabs.harvard.edu/abs/2026MNRAS.545f2015C} {545, staf2015}

\bibitem[\protect\citeauthoryear{{Concha-Ram{\'\i}rez}, {Wilhelm}, {Portegies Zwart}  \& {Haworth}}{{Concha-Ram{\'\i}rez} et~al.}{2019}]{ConchaRamirez19}
{Concha-Ram{\'\i}rez} F.,  {Wilhelm} M. J.~C.,  {Portegies Zwart} S.,   {Haworth} T.~J.,  2019, \mn@doi [\mnras] {10.1093/mnras/stz2973}, \href {https://ui.adsabs.harvard.edu/abs/2019MNRAS.490.5678C} {490, 5678}

\bibitem[\protect\citeauthoryear{{D'Angelo} \& {Marzari}}{{D'Angelo} \& {Marzari}}{2012}]{Dangelo12}
{D'Angelo} G.,  {Marzari} F.,  2012, \mn@doi [\apj] {10.1088/0004-637X/757/1/50}, \href {http://adsabs.harvard.edu/abs/2012ApJ...757...50D} {757, 50}

\bibitem[\protect\citeauthoryear{{Dittrich}, {Klahr}  \& {Johansen}}{{Dittrich} et~al.}{2013}]{Dittrich13}
{Dittrich} K.,  {Klahr} H.,   {Johansen} A.,  2013, \mn@doi [\apj] {10.1088/0004-637X/763/2/117}, \href {http://adsabs.harvard.edu/abs/2013ApJ...763..117D} {763, 117}

\bibitem[\protect\citeauthoryear{{Doyle} et~al.,}{{Doyle} et~al.}{2011}]{Doyle11}
{Doyle} L.~R.,  et~al., 2011, \mn@doi [Science] {10.1126/science.1210923}, \href {https://ui.adsabs.harvard.edu/abs/2011Sci...333.1602D} {333, 1602}

\bibitem[\protect\citeauthoryear{{Eggenberger}, {Udry}, {Chauvin}, {Beuzit}, {Lagrange}, {S{\'e}gransan}  \& {Mayor}}{{Eggenberger} et~al.}{2007}]{Eggenberger07}
{Eggenberger} A.,  {Udry} S.,  {Chauvin} G.,  {Beuzit} J.~L.,  {Lagrange} A.~M.,  {S{\'e}gransan} D.,   {Mayor} M.,  2007, \mn@doi [\aap] {10.1051/0004-6361:20077447}, \href {https://ui.adsabs.harvard.edu/abs/2007A&A...474..273E} {474, 273}

\bibitem[\protect\citeauthoryear{{Eisner} et~al.,}{{Eisner} et~al.}{2018}]{Eisner18}
{Eisner} J.~A.,  et~al., 2018, \mn@doi [\apj] {10.3847/1538-4357/aac3e2}, \href {https://ui.adsabs.harvard.edu/abs/2018ApJ...860...77E} {860, 77}

\bibitem[\protect\citeauthoryear{{Emsenhuber}, {Mordasini}, {Burn}, {Alibert}, {Benz}  \& {Asphaug}}{{Emsenhuber} et~al.}{2021}]{Emsenhuber21a}
{Emsenhuber} A.,  {Mordasini} C.,  {Burn} R.,  {Alibert} Y.,  {Benz} W.,   {Asphaug} E.,  2021, \mn@doi [\aap] {10.1051/0004-6361/202038553}, \href {https://ui.adsabs.harvard.edu/abs/2021A&A...656A..69E} {656, A69}

\bibitem[\protect\citeauthoryear{{Emsenhuber} et~al.,}{{Emsenhuber} et~al.}{2025}]{Emsenhuber25}
{Emsenhuber} A.,  et~al., 2025, \mn@doi [\aap] {10.1051/0004-6361/202452485}, \href {https://ui.adsabs.harvard.edu/abs/2025A&A...701A..64E} {701, A64}

\bibitem[\protect\citeauthoryear{{Ercolano}, {Picogna}, {Monsch}, {Drake}  \& {Preibisch}}{{Ercolano} et~al.}{2021}]{Ercolano21}
{Ercolano} B.,  {Picogna} G.,  {Monsch} K.,  {Drake} J.~J.,   {Preibisch} T.,  2021, \mn@doi [\mnras] {10.1093/mnras/stab2590}, \href {https://ui.adsabs.harvard.edu/abs/2021MNRAS.508.1675E} {508, 1675}

\bibitem[\protect\citeauthoryear{{Flock}, {Nelson}, {Turner}, {Bertrang}, {Carrasco-Gonz{\'a}lez}, {Henning}, {Lyra}  \& {Teague}}{{Flock} et~al.}{2017}]{Flock17}
{Flock} M.,  {Nelson} R.~P.,  {Turner} N.~J.,  {Bertrang} G. H.~M.,  {Carrasco-Gonz{\'a}lez} C.,  {Henning} T.,  {Lyra} W.,   {Teague} R.,  2017, \mn@doi [\apj] {10.3847/1538-4357/aa943f}, \href {https://ui.adsabs.harvard.edu/abs/2017ApJ...850..131F} {850, 131}

\bibitem[\protect\citeauthoryear{{Flock}, {Turner}, {Nelson}, {Lyra}, {Manger}  \& {Klahr}}{{Flock} et~al.}{2020}]{Flock20}
{Flock} M.,  {Turner} N.~J.,  {Nelson} R.~P.,  {Lyra} W.,  {Manger} N.,   {Klahr} H.,  2020, \mn@doi [\apj] {10.3847/1538-4357/ab9641}, \href {https://ui.adsabs.harvard.edu/abs/2020ApJ...897..155F} {897, 155}

\bibitem[\protect\citeauthoryear{{Fortier}, {Alibert}, {Carron}, {Benz}  \& {Dittkrist}}{{Fortier} et~al.}{2013}]{Fortier13}
{Fortier} A.,  {Alibert} Y.,  {Carron} F.,  {Benz} W.,   {Dittkrist} K.~M.,  2013, \mn@doi [\aap] {10.1051/0004-6361/201220241}, \href {https://ui.adsabs.harvard.edu/abs/2013A&A...549A..44F} {549, A44}

\bibitem[\protect\citeauthoryear{{Fromang} \& {Nelson}}{{Fromang} \& {Nelson}}{2006}]{FromangNelson2006}
{Fromang} S.,  {Nelson} R.~P.,  2006, \mn@doi [\aap] {10.1051/0004-6361:20065643}, \href {http://adsabs.harvard.edu/abs/2006A%26A...457..343F} {457, 343}

\bibitem[\protect\citeauthoryear{{Goldreich} \& {Ward}}{{Goldreich} \& {Ward}}{1973}]{GoldreichWard73}
{Goldreich} P.,  {Ward} W.~R.,  1973, \mn@doi [\apj] {10.1086/152291}, \href {https://ui.adsabs.harvard.edu/abs/1973ApJ...183.1051G} {183, 1051}

\bibitem[\protect\citeauthoryear{{Gorti}, {Dullemond}  \& {Hollenbach}}{{Gorti} et~al.}{2009}]{Gorti09}
{Gorti} U.,  {Dullemond} C.~P.,   {Hollenbach} D.,  2009, \mn@doi [\apj] {10.1088/0004-637X/705/2/1237}, \href {https://ui.adsabs.harvard.edu/abs/2009ApJ...705.1237G} {705, 1237}

\bibitem[\protect\citeauthoryear{{Gorti}, {Hollenbach}  \& {Dullemond}}{{Gorti} et~al.}{2015}]{Gorti15}
{Gorti} U.,  {Hollenbach} D.,   {Dullemond} C.~P.,  2015, \mn@doi [\apj] {10.1088/0004-637X/804/1/29}, \href {https://ui.adsabs.harvard.edu/abs/2015ApJ...804...29G} {804, 29}

\bibitem[\protect\citeauthoryear{{Guarcello} et~al.,}{{Guarcello} et~al.}{2016}]{Guarcello16}
{Guarcello} M.~G.,  et~al., 2016, \mn@doi [arXiv e-prints] {10.48550/arXiv.1605.01773}, \href {https://ui.adsabs.harvard.edu/abs/2016arXiv160501773G} {p. arXiv:1605.01773}

\bibitem[\protect\citeauthoryear{{Harris}, {Andrews}, {Wilner}  \& {Kraus}}{{Harris} et~al.}{2012}]{Harris12}
{Harris} R.~J.,  {Andrews} S.~M.,  {Wilner} D.~J.,   {Kraus} A.~L.,  2012, \mn@doi [\apj] {10.1088/0004-637X/751/2/115}, \href {https://ui.adsabs.harvard.edu/abs/2012ApJ...751..115H} {751, 115}

\bibitem[\protect\citeauthoryear{{Hatzes}, {Cochran}, {Endl}, {McArthur}, {Paulson}, {Walker}, {Campbell}  \& {Yang}}{{Hatzes} et~al.}{2003}]{Hatzes03}
{Hatzes} A.~P.,  {Cochran} W.~D.,  {Endl} M.,  {McArthur} B.,  {Paulson} D.~B.,  {Walker} G. A.~H.,  {Campbell} B.,   {Yang} S.,  2003, \mn@doi [\apj] {10.1086/379281}, \href {https://ui.adsabs.harvard.edu/abs/2003ApJ...599.1383H} {599, 1383}

\bibitem[\protect\citeauthoryear{{Haworth} \& {Clarke}}{{Haworth} \& {Clarke}}{2019}]{Haworth19}
{Haworth} T.~J.,  {Clarke} C.~J.,  2019, \mn@doi [\mnras] {10.1093/mnras/stz706}, \href {https://ui.adsabs.harvard.edu/abs/2019MNRAS.485.3895H} {485, 3895}

\bibitem[\protect\citeauthoryear{{Haworth}, {Facchini}, {Clarke}  \& {Cleeves}}{{Haworth} et~al.}{2017}]{Haworth17}
{Haworth} T.~J.,  {Facchini} S.,  {Clarke} C.~J.,   {Cleeves} L.~I.,  2017, \mn@doi [\mnras] {10.1093/mnrasl/slx037}, \href {https://ui.adsabs.harvard.edu/abs/2017MNRAS.468L.108H} {468, L108}

\bibitem[\protect\citeauthoryear{{Haworth}, {Clarke}, {Rahman}, {Winter}  \& {Facchini}}{{Haworth} et~al.}{2018}]{Haworth18}
{Haworth} T.~J.,  {Clarke} C.~J.,  {Rahman} W.,  {Winter} A.~J.,   {Facchini} S.,  2018, \mn@doi [\mnras] {10.1093/mnras/sty2323}, \href {https://ui.adsabs.harvard.edu/abs/2018MNRAS.481..452H} {481, 452}

\bibitem[\protect\citeauthoryear{{Haworth}, {Cadman}, {Meru}, {Hall}, {Albertini}, {Forgan}, {Rice}  \& {Owen}}{{Haworth} et~al.}{2020}]{Haworth20}
{Haworth} T.~J.,  {Cadman} J.,  {Meru} F.,  {Hall} C.,  {Albertini} E.,  {Forgan} D.,  {Rice} K.,   {Owen} J.~E.,  2020, \mn@doi [\mnras] {10.1093/mnras/staa883}, \href {https://ui.adsabs.harvard.edu/abs/2020MNRAS.494.4130H} {494, 4130}

\bibitem[\protect\citeauthoryear{{Haworth}, {Coleman}, {Qiao}, {Sellek}  \& {Askari}}{{Haworth} et~al.}{2023}]{Haworth23}
{Haworth} T.~J.,  {Coleman} G. A.~L.,  {Qiao} L.,  {Sellek} A.~D.,   {Askari} K.,  2023, \mn@doi [\mnras] {10.1093/mnras/stad3054}, \href {https://ui.adsabs.harvard.edu/abs/2023MNRAS.526.4315H} {526, 4315}

\bibitem[\protect\citeauthoryear{{Hirsch} et~al.,}{{Hirsch} et~al.}{2021}]{Hirsch21}
{Hirsch} L.~A.,  et~al., 2021, \mn@doi [\aj] {10.3847/1538-3881/abd639}, \href {https://ui.adsabs.harvard.edu/abs/2021AJ....161..134H} {161, 134}

\bibitem[\protect\citeauthoryear{{H{\"u}hn} et~al.,}{{H{\"u}hn} et~al.}{2025}]{Huhn25}
{H{\"u}hn} L.-A.,  et~al., 2025, \mn@doi [\aap] {10.1051/0004-6361/202452689}, \href {https://ui.adsabs.harvard.edu/abs/2025A&A...696A.162H} {696, A162}

\bibitem[\protect\citeauthoryear{{Ida}, {Lin}  \& {Nagasawa}}{{Ida} et~al.}{2013}]{Ida13}
{Ida} S.,  {Lin} D.~N.~C.,   {Nagasawa} M.,  2013, \mn@doi [\apj] {10.1088/0004-637X/775/1/42}, \href {http://adsabs.harvard.edu/abs/2013ApJ...775...42I} {775, 42}

\bibitem[\protect\citeauthoryear{{Inaba}, {Tanaka}, {Nakazawa}, {Wetherill}  \& {Kokubo}}{{Inaba} et~al.}{2001}]{Inaba01}
{Inaba} S.,  {Tanaka} H.,  {Nakazawa} K.,  {Wetherill} G.~W.,   {Kokubo} E.,  2001, \mn@doi [\icarus] {10.1006/icar.2000.6533}, \href {https://ui.adsabs.harvard.edu/abs/2001Icar..149..235I} {149, 235}

\bibitem[\protect\citeauthoryear{{Jang-Condell}, {Mugrauer}  \& {Schmidt}}{{Jang-Condell} et~al.}{2008}]{Jang-Condell08}
{Jang-Condell} H.,  {Mugrauer} M.,   {Schmidt} T.,  2008, \mn@doi [\apjl] {10.1086/591791}, \href {https://ui.adsabs.harvard.edu/abs/2008ApJ...683L.191J} {683, L191}

\bibitem[\protect\citeauthoryear{{Johansen} \& {Lambrechts}}{{Johansen} \& {Lambrechts}}{2017}]{Johansen17}
{Johansen} A.,  {Lambrechts} M.,  2017, \mn@doi [Annual Review of Earth and Planetary Sciences] {10.1146/annurev-earth-063016-020226}, \href {http://adsabs.harvard.edu/abs/2017AREPS..45..359J} {45, 359}

\bibitem[\protect\citeauthoryear{{Johansen}, {Oishi}, {Mac Low}, {Klahr}, {Henning}  \& {Youdin}}{{Johansen} et~al.}{2007}]{Johansen07}
{Johansen} A.,  {Oishi} J.~S.,  {Mac Low} M.-M.,  {Klahr} H.,  {Henning} T.,   {Youdin} A.,  2007, \mn@doi [\nat] {10.1038/nature06086}, \href {http://adsabs.harvard.edu/abs/2007Natur.448.1022J} {448, 1022}

\bibitem[\protect\citeauthoryear{{Johansen}, {Youdin}  \& {Klahr}}{{Johansen} et~al.}{2009a}]{JohansenYoudin2009}
{Johansen} A.,  {Youdin} A.,   {Klahr} H.,  2009a, \mn@doi [\apj] {10.1088/0004-637X/697/2/1269}, \href {http://adsabs.harvard.edu/abs/2009ApJ...697.1269J} {697, 1269}

\bibitem[\protect\citeauthoryear{{Johansen}, {Youdin}  \& {Mac Low}}{{Johansen} et~al.}{2009b}]{Johansen09}
{Johansen} A.,  {Youdin} A.,   {Mac Low} M.-M.,  2009b, \mn@doi [\apjl] {10.1088/0004-637X/704/2/L75}, \href {https://ui.adsabs.harvard.edu/abs/2009ApJ...704L..75J} {704, L75}

\bibitem[\protect\citeauthoryear{{Johansen}, {Blum}, {Tanaka}, {Ormel}, {Bizzarro}  \& {Rickman}}{{Johansen} et~al.}{2014}]{Johansen14}
{Johansen} A.,  {Blum} J.,  {Tanaka} H.,  {Ormel} C.,  {Bizzarro} M.,   {Rickman} H.,  2014, in {Beuther} H.,  {Klessen} R.~S.,  {Dullemond} C.~P.,   {Henning} T.,  eds, Protostars and Planets VI. p.~547 (\mn@eprint {arXiv} {1402.1344}), \mn@doi{10.2458/azu_uapress_9780816531240-ch024}

\bibitem[\protect\citeauthoryear{{Johansen}, {Mac Low}, {Lacerda}  \& {Bizzarro}}{{Johansen} et~al.}{2015}]{Johansen15}
{Johansen} A.,  {Mac Low} M.-M.,  {Lacerda} P.,   {Bizzarro} M.,  2015, \mn@doi [Science Advances] {10.1126/sciadv.1500109}, \href {https://ui.adsabs.harvard.edu/abs/2015SciA....1E0109J} {1, 1500109}

\bibitem[\protect\citeauthoryear{{Kley} \& {Nelson}}{{Kley} \& {Nelson}}{2008}]{Kley08}
{Kley} W.,  {Nelson} R.~P.,  2008, \mn@doi [\aap] {10.1051/0004-6361:20079324}, \href {https://ui.adsabs.harvard.edu/abs/2008A&A...486..617K} {486, 617}

\bibitem[\protect\citeauthoryear{{Komaki}, {Fukuhara}, {Suzuki}  \& {Yoshida}}{{Komaki} et~al.}{2023}]{Komaki23}
{Komaki} A.,  {Fukuhara} S.,  {Suzuki} T.~K.,   {Yoshida} N.,  2023, \mn@doi [arXiv e-prints] {10.48550/arXiv.2304.13316}, \href {https://ui.adsabs.harvard.edu/abs/2023arXiv230413316K} {p. arXiv:2304.13316}

\bibitem[\protect\citeauthoryear{{Kraus}, {Ireland}, {Hillenbrand}  \& {Martinache}}{{Kraus} et~al.}{2012}]{Kraus12}
{Kraus} A.~L.,  {Ireland} M.~J.,  {Hillenbrand} L.~A.,   {Martinache} F.,  2012, \mn@doi [\apj] {10.1088/0004-637X/745/1/19}, \href {https://ui.adsabs.harvard.edu/abs/2012ApJ...745...19K} {745, 19}

\bibitem[\protect\citeauthoryear{{Lambrechts} \& {Johansen}}{{Lambrechts} \& {Johansen}}{2012}]{Lambrechts12}
{Lambrechts} M.,  {Johansen} A.,  2012, \mn@doi [\aap] {10.1051/0004-6361/201219127}, \href {http://adsabs.harvard.edu/abs/2012A%26A...544A..32L} {544, A32}

\bibitem[\protect\citeauthoryear{{Lambrechts} \& {Johansen}}{{Lambrechts} \& {Johansen}}{2014}]{Lambrechts14}
{Lambrechts} M.,  {Johansen} A.,  2014, \mn@doi [\aap] {10.1051/0004-6361/201424343}, \href {http://adsabs.harvard.edu/abs/2014A%26A...572A.107L} {572, A107}

\bibitem[\protect\citeauthoryear{{Latter}}{{Latter}}{2016}]{Latter16}
{Latter} H.~N.,  2016, \mn@doi [\mnras] {10.1093/mnras/stv2449}, \href {https://ui.adsabs.harvard.edu/abs/2016MNRAS.455.2608L} {455, 2608}

\bibitem[\protect\citeauthoryear{{Lenz}, {Klahr}  \& {Birnstiel}}{{Lenz} et~al.}{2019}]{Lenz19}
{Lenz} C.~T.,  {Klahr} H.,   {Birnstiel} T.,  2019, \mn@doi [\apj] {10.3847/1538-4357/ab05d9}, \href {https://ui.adsabs.harvard.edu/abs/2019ApJ...874...36L} {874, 36}

\bibitem[\protect\citeauthoryear{{Lesur} et~al.,}{{Lesur} et~al.}{2023}]{Lesur23}
{Lesur} G.,  et~al., 2023, in {Inutsuka} S.,  {Aikawa} Y.,  {Muto} T.,  {Tomida} K.,   {Tamura} M.,  eds,  Astronomical Society of the Pacific Conference Series Vol. 534, Protostars and Planets VII. p.~465 (\mn@eprint {arXiv} {2203.09821}), \mn@doi{10.48550/arXiv.2203.09821}

\bibitem[\protect\citeauthoryear{{Li}, {Youdin}  \& {Simon}}{{Li} et~al.}{2019}]{Li19}
{Li} R.,  {Youdin} A.~N.,   {Simon} J.~B.,  2019, \mn@doi [\apj] {10.3847/1538-4357/ab480d}, \href {https://ui.adsabs.harvard.edu/abs/2019ApJ...885...69L} {885, 69}

\bibitem[\protect\citeauthoryear{{Lin} \& {Papaloizou}}{{Lin} \& {Papaloizou}}{1986}]{LinPapaloizou86}
{Lin} D.~N.~C.,  {Papaloizou} J.,  1986, \mn@doi [\apj] {10.1086/164653}, \href {http://adsabs.harvard.edu/abs/1986ApJ...309..846L} {309, 846}

\bibitem[\protect\citeauthoryear{{Liu}, {Lambrechts}, {Johansen}, {Pascucci}  \& {Henning}}{{Liu} et~al.}{2020}]{Liu20}
{Liu} B.,  {Lambrechts} M.,  {Johansen} A.,  {Pascucci} I.,   {Henning} T.,  2020, \mn@doi [\aap] {10.1051/0004-6361/202037720}, \href {https://ui.adsabs.harvard.edu/abs/2020A&A...638A..88L} {638, A88}

\bibitem[\protect\citeauthoryear{{Lynden-Bell} \& {Pringle}}{{Lynden-Bell} \& {Pringle}}{1974}]{Lynden-BellPringle1974}
{Lynden-Bell} D.,  {Pringle} J.~E.,  1974, \mnras, \href {http://adsabs.harvard.edu/abs/1974MNRAS.168..603L} {168, 603}

\bibitem[\protect\citeauthoryear{{Mann} et~al.,}{{Mann} et~al.}{2014}]{Mann14}
{Mann} R.~K.,  et~al., 2014, \mn@doi [\apj] {10.1088/0004-637X/784/1/82}, \href {https://ui.adsabs.harvard.edu/abs/2014ApJ...784...82M} {784, 82}

\bibitem[\protect\citeauthoryear{{Marzari} \& {Scholl}}{{Marzari} \& {Scholl}}{2000}]{Marzari00}
{Marzari} F.,  {Scholl} H.,  2000, \mn@doi [\apj] {10.1086/317091}, \href {https://ui.adsabs.harvard.edu/abs/2000ApJ...543..328M} {543, 328}

\bibitem[\protect\citeauthoryear{{Matsuyama}, {Johnstone}  \& {Hartmann}}{{Matsuyama} et~al.}{2003}]{Matsuyama03}
{Matsuyama} I.,  {Johnstone} D.,   {Hartmann} L.,  2003, \mn@doi [\apj] {10.1086/344638}, \href {https://ui.adsabs.harvard.edu/abs/2003ApJ...582..893M} {582, 893}

\bibitem[\protect\citeauthoryear{Meerschaert, Roy  \& Shao}{Meerschaert et~al.}{2012}]{Meerschaert12}
Meerschaert M.~M.,  Roy P.,   Shao Q.,  2012, \mn@doi [Communications in Statistics - Theory and Methods] {10.1080/03610926.2011.552828}, 41, 1839

\bibitem[\protect\citeauthoryear{{Mordasini}, {Molli{\`e}re}, {Dittkrist}, {Jin}  \& {Alibert}}{{Mordasini} et~al.}{2015}]{Mordasini15}
{Mordasini} C.,  {Molli{\`e}re} P.,  {Dittkrist} K.-M.,  {Jin} S.,   {Alibert} Y.,  2015, \mn@doi [International Journal of Astrobiology] {10.1017/S1473550414000263}, \href {http://adsabs.harvard.edu/abs/2015IJAsB..14..201M} {14, 201}

\bibitem[\protect\citeauthoryear{{Nakatani}, {Hosokawa}, {Yoshida}, {Nomura}  \& {Kuiper}}{{Nakatani} et~al.}{2018}]{Nakatani18}
{Nakatani} R.,  {Hosokawa} T.,  {Yoshida} N.,  {Nomura} H.,   {Kuiper} R.,  2018, \mn@doi [\apj] {10.3847/1538-4357/aad9fd}, \href {https://ui.adsabs.harvard.edu/abs/2018ApJ...865...75N} {865, 75}

\bibitem[\protect\citeauthoryear{{Nelson}, {Gressel}  \& {Umurhan}}{{Nelson} et~al.}{2013}]{Nelson13}
{Nelson} R.~P.,  {Gressel} O.,   {Umurhan} O.~M.,  2013, \mn@doi [\mnras] {10.1093/mnras/stt1475}, \href {https://ui.adsabs.harvard.edu/abs/2013MNRAS.435.2610N} {435, 2610}

\bibitem[\protect\citeauthoryear{{O'Dell} \& {Wen}}{{O'Dell} \& {Wen}}{1994}]{Odell94}
{O'Dell} C.~R.,  {Wen} Z.,  1994, \mn@doi [\apj] {10.1086/174892}, \href {https://ui.adsabs.harvard.edu/abs/1994ApJ...436..194O} {436, 194}

\bibitem[\protect\citeauthoryear{{Offner}, {Moe}, {Kratter}, {Sadavoy}, {Jensen}  \& {Tobin}}{{Offner} et~al.}{2023}]{Offner23}
{Offner} S.~S.~R.,  {Moe} M.,  {Kratter} K.~M.,  {Sadavoy} S.~I.,  {Jensen} E.~L.~N.,   {Tobin} J.~J.,  2023, in {Inutsuka} S.,  {Aikawa} Y.,  {Muto} T.,  {Tomida} K.,   {Tamura} M.,  eds,  Astronomical Society of the Pacific Conference Series Vol. 534, Protostars and Planets VII. p.~275 (\mn@eprint {arXiv} {2203.10066}), \mn@doi{10.48550/arXiv.2203.10066}

\bibitem[\protect\citeauthoryear{{Ohtsuki}}{{Ohtsuki}}{1999}]{Ohtsuki99}
{Ohtsuki} K.,  1999, \mn@doi [\icarus] {10.1006/icar.1998.6041}, \href {https://ui.adsabs.harvard.edu/abs/1999Icar..137..152O} {137, 152}

\bibitem[\protect\citeauthoryear{{Ormel} \& {Klahr}}{{Ormel} \& {Klahr}}{2010}]{OrmelKlahr2010}
{Ormel} C.~W.,  {Klahr} H.~H.,  2010, \mn@doi [\aap] {10.1051/0004-6361/201014903}, \href {http://adsabs.harvard.edu/abs/2010A%26A...520A..43O} {520, A43}

\bibitem[\protect\citeauthoryear{{Owen}, {Ercolano}, {Clarke}  \& {Alexander}}{{Owen} et~al.}{2010}]{Owen10}
{Owen} J.~E.,  {Ercolano} B.,  {Clarke} C.~J.,   {Alexander} R.~D.,  2010, \mn@doi [\mnras] {10.1111/j.1365-2966.2009.15771.x}, \href {https://ui.adsabs.harvard.edu/abs/2010MNRAS.401.1415O} {401, 1415}

\bibitem[\protect\citeauthoryear{{Owen}, {Clarke}  \& {Ercolano}}{{Owen} et~al.}{2012}]{Owen12}
{Owen} J.~E.,  {Clarke} C.~J.,   {Ercolano} B.,  2012, \mn@doi [\mnras] {10.1111/j.1365-2966.2011.20337.x}, \href {https://ui.adsabs.harvard.edu/abs/2012MNRAS.422.1880O} {422, 1880}

\bibitem[\protect\citeauthoryear{{Paardekooper} \& {Mellema}}{{Paardekooper} \& {Mellema}}{2006}]{PaardekooperMellema06}
{Paardekooper} S.-J.,  {Mellema} G.,  2006, \mn@doi [\aap] {10.1051/0004-6361:20066304}, \href {http://adsabs.harvard.edu/abs/2006A%26A...459L..17P} {459, L17}

\bibitem[\protect\citeauthoryear{{Paardekooper}, {Th{\'e}bault}  \& {Mellema}}{{Paardekooper} et~al.}{2008}]{Paardekooper08}
{Paardekooper} S.~J.,  {Th{\'e}bault} P.,   {Mellema} G.,  2008, \mn@doi [\mnras] {10.1111/j.1365-2966.2008.13080.x}, \href {https://ui.adsabs.harvard.edu/abs/2008MNRAS.386..973P} {386, 973}

\bibitem[\protect\citeauthoryear{{Paardekooper}, {Baruteau}, {Crida}  \& {Kley}}{{Paardekooper} et~al.}{2010}]{pdk10}
{Paardekooper} S.-J.,  {Baruteau} C.,  {Crida} A.,   {Kley} W.,  2010, \mn@doi [\mnras] {10.1111/j.1365-2966.2009.15782.x}, \href {http://adsabs.harvard.edu/abs/2010MNRAS.401.1950P} {401, 1950}

\bibitem[\protect\citeauthoryear{{Paardekooper}, {Baruteau}  \& {Kley}}{{Paardekooper} et~al.}{2011}]{pdk11}
{Paardekooper} S.-J.,  {Baruteau} C.,   {Kley} W.,  2011, \mn@doi [\mnras] {10.1111/j.1365-2966.2010.17442.x}, \href {http://adsabs.harvard.edu/abs/2011MNRAS.410..293P} {410, 293}

\bibitem[\protect\citeauthoryear{{Papaloizou} \& {Nelson}}{{Papaloizou} \& {Nelson}}{2003}]{PapaloizouNelson2003}
{Papaloizou} J.~C.~B.,  {Nelson} R.~P.,  2003, \mn@doi [\mnras] {10.1046/j.1365-8711.2003.06246.x}, \href {http://adsabs.harvard.edu/abs/2003MNRAS.339..983P} {339, 983}

\bibitem[\protect\citeauthoryear{{Papaloizou} \& {Pringle}}{{Papaloizou} \& {Pringle}}{1977}]{Papaloizou77}
{Papaloizou} J.,  {Pringle} J.~E.,  1977, \mn@doi [\mnras] {10.1093/mnras/181.3.441}, \href {https://ui.adsabs.harvard.edu/abs/1977MNRAS.181..441P} {181, 441}

\bibitem[\protect\citeauthoryear{{Picogna}, {Ercolano}, {Owen}  \& {Weber}}{{Picogna} et~al.}{2019}]{Picogna19}
{Picogna} G.,  {Ercolano} B.,  {Owen} J.~E.,   {Weber} M.~L.,  2019, \mn@doi [\mnras] {10.1093/mnras/stz1166}, \href {https://ui.adsabs.harvard.edu/abs/2019MNRAS.487..691P} {487, 691}

\bibitem[\protect\citeauthoryear{{Picogna}, {Ercolano}  \& {Espaillat}}{{Picogna} et~al.}{2021}]{Picogna21}
{Picogna} G.,  {Ercolano} B.,   {Espaillat} C.~C.,  2021, \mn@doi [\mnras] {10.1093/mnras/stab2883}, \href {https://ui.adsabs.harvard.edu/abs/2021MNRAS.508.3611P} {508, 3611}

\bibitem[\protect\citeauthoryear{{Poon}, {Nelson}  \& {Coleman}}{{Poon} et~al.}{2021}]{Poon21}
{Poon} S. T.~S.,  {Nelson} R.~P.,   {Coleman} G. A.~L.,  2021, \mn@doi [\mnras] {10.1093/mnras/stab1466}, \href {https://ui.adsabs.harvard.edu/abs/2021MNRAS.505.2500P} {505, 2500}

\bibitem[\protect\citeauthoryear{{Qiao}, {Haworth}, {Sellek}  \& {Ali}}{{Qiao} et~al.}{2022}]{Qiao22}
{Qiao} L.,  {Haworth} T.~J.,  {Sellek} A.~D.,   {Ali} A.~A.,  2022, \mn@doi [\mnras] {10.1093/mnras/stac684}, \href {https://ui.adsabs.harvard.edu/abs/2022MNRAS.512.3788Q} {512, 3788}

\bibitem[\protect\citeauthoryear{{Qiao}, {Coleman}  \& {Haworth}}{{Qiao} et~al.}{2023}]{Qiao23}
{Qiao} L.,  {Coleman} G. A.~L.,   {Haworth} T.~J.,  2023, \mn@doi [\mnras] {10.1093/mnras/stad944}, \href {https://ui.adsabs.harvard.edu/abs/2023MNRAS.522.1939Q} {522, 1939}

\bibitem[\protect\citeauthoryear{{Qiao}, {Coleman}  \& {Haworth}}{{Qiao} et~al.}{2026}]{Qiao26}
{Qiao} L.,  {Coleman} G. A.~L.,   {Haworth} T.~J.,  2026, \mn@doi [\mnras] {10.1093/mnras/stag034}, \href {https://ui.adsabs.harvard.edu/abs/2026MNRAS.546ag034Q} {546, stag034}

\bibitem[\protect\citeauthoryear{{Quintana}, {Lissauer}, {Chambers}  \& {Duncan}}{{Quintana} et~al.}{2002}]{Quintana02}
{Quintana} E.~V.,  {Lissauer} J.~J.,  {Chambers} J.~E.,   {Duncan} M.~J.,  2002, \mn@doi [\apj] {10.1086/341808}, \href {https://ui.adsabs.harvard.edu/abs/2002ApJ...576..982Q} {576, 982}

\bibitem[\protect\citeauthoryear{{Quintana}, {Adams}, {Lissauer}  \& {Chambers}}{{Quintana} et~al.}{2007}]{Quintana07}
{Quintana} E.~V.,  {Adams} F.~C.,  {Lissauer} J.~J.,   {Chambers} J.~E.,  2007, \mn@doi [\apj] {10.1086/512542}, \href {https://ui.adsabs.harvard.edu/abs/2007ApJ...660..807Q} {660, 807}

\bibitem[\protect\citeauthoryear{{Rafikov}}{{Rafikov}}{2004}]{Rafikov04}
{Rafikov} R.~R.,  2004, \mn@doi [\aj] {10.1086/423216}, \href {https://ui.adsabs.harvard.edu/abs/2004AJ....128.1348R} {128, 1348}

\bibitem[\protect\citeauthoryear{{Raghavan} et~al.,}{{Raghavan} et~al.}{2010}]{Raghavan10}
{Raghavan} D.,  et~al., 2010, \mn@doi [\apjs] {10.1088/0067-0049/190/1/1}, \href {https://ui.adsabs.harvard.edu/abs/2010ApJS..190....1R} {190, 1}

\bibitem[\protect\citeauthoryear{{Rice}, {Armitage}, {Wood}  \& {Lodato}}{{Rice} et~al.}{2006}]{Rice06}
{Rice} W.~K.~M.,  {Armitage} P.~J.,  {Wood} K.,   {Lodato} G.,  2006, \mn@doi [\mnras] {10.1111/j.1365-2966.2006.11113.x}, \href {https://ui.adsabs.harvard.edu/abs/2006MNRAS.373.1619R} {373, 1619}

\bibitem[\protect\citeauthoryear{{Roell}, {Neuh{\"a}user}, {Seifahrt}  \& {Mugrauer}}{{Roell} et~al.}{2012}]{Roell12}
{Roell} T.,  {Neuh{\"a}user} R.,  {Seifahrt} A.,   {Mugrauer} M.,  2012, \mn@doi [\aap] {10.1051/0004-6361/201118051}, \href {https://ui.adsabs.harvard.edu/abs/2012A&A...542A..92R} {542, A92}

\bibitem[\protect\citeauthoryear{{Rosotti} \& {Clarke}}{{Rosotti} \& {Clarke}}{2018}]{Rosotti18}
{Rosotti} G.~P.,  {Clarke} C.~J.,  2018, \mn@doi [\mnras] {10.1093/mnras/stx2769}, \href {https://ui.adsabs.harvard.edu/abs/2018MNRAS.473.5630R} {473, 5630}

\bibitem[\protect\citeauthoryear{{Rosotti}, {Teague}, {Dullemond}, {Booth}  \& {Clarke}}{{Rosotti} et~al.}{2020}]{Rosotti20}
{Rosotti} G.~P.,  {Teague} R.,  {Dullemond} C.,  {Booth} R.~A.,   {Clarke} C.~J.,  2020, \mn@doi [\mnras] {10.1093/mnras/staa1170}, \href {https://ui.adsabs.harvard.edu/abs/2020MNRAS.495..173R} {495, 173}

\bibitem[\protect\citeauthoryear{{Sairam} et~al.,}{{Sairam} et~al.}{2024}]{Sairam24}
{Sairam} L.,  et~al., 2024, \mn@doi [\mnras] {10.1093/mnras/stae2317}, \href {https://ui.adsabs.harvard.edu/abs/2024MNRAS.534.3999S} {534, 3999}

\bibitem[\protect\citeauthoryear{{Savonije}, {Papaloizou}  \& {Lin}}{{Savonije} et~al.}{1994}]{Savonije94}
{Savonije} G.~J.,  {Papaloizou} J.~C.~B.,   {Lin} D.~N.~C.,  1994, \mn@doi [\mnras] {10.1093/mnras/268.1.13}, \href {https://ui.adsabs.harvard.edu/abs/1994MNRAS.268...13S} {268, 13}

\bibitem[\protect\citeauthoryear{{Savvidou} \& {Bitsch}}{{Savvidou} \& {Bitsch}}{2023}]{Savvidou23}
{Savvidou} S.,  {Bitsch} B.,  2023, \mn@doi [\aap] {10.1051/0004-6361/202245793}, \href {https://ui.adsabs.harvard.edu/abs/2023A&A...679A..42S} {679, A42}

\bibitem[\protect\citeauthoryear{{Sch{\"a}fer}, {Yang}  \& {Johansen}}{{Sch{\"a}fer} et~al.}{2017}]{Schafer17}
{Sch{\"a}fer} U.,  {Yang} C.-C.,   {Johansen} A.,  2017, \mn@doi [\aap] {10.1051/0004-6361/201629561}, \href {https://ui.adsabs.harvard.edu/abs/2017A&A...597A..69S} {597, A69}

\bibitem[\protect\citeauthoryear{{Schreiber} \& {Klahr}}{{Schreiber} \& {Klahr}}{2018}]{Schreiber18}
{Schreiber} A.,  {Klahr} H.,  2018, \mn@doi [\apj] {10.3847/1538-4357/aac3d4}, \href {https://ui.adsabs.harvard.edu/abs/2018ApJ...861...47S} {861, 47}

\bibitem[\protect\citeauthoryear{{Schroetter} et~al.,}{{Schroetter} et~al.}{2025}]{Schroetter25}
{Schroetter} I.,  et~al., 2025, \mn@doi [Nature Astronomy] {10.1038/s41550-025-02596-6}, \href {https://ui.adsabs.harvard.edu/abs/2025NatAs...9.1326S} {9, 1326}

\bibitem[\protect\citeauthoryear{{Sellek}, {Booth}  \& {Clarke}}{{Sellek} et~al.}{2020}]{Sellek20}
{Sellek} A.~D.,  {Booth} R.~A.,   {Clarke} C.~J.,  2020, \mn@doi [\mnras] {10.1093/mnras/stz3528}, \href {https://ui.adsabs.harvard.edu/abs/2020MNRAS.492.1279S} {492, 1279}

\bibitem[\protect\citeauthoryear{{Sellek}, {Clarke}  \& {Ercolano}}{{Sellek} et~al.}{2022}]{Sellek22}
{Sellek} A.~D.,  {Clarke} C.~J.,   {Ercolano} B.,  2022, \mn@doi [\mnras] {10.1093/mnras/stac1148}, \href {https://ui.adsabs.harvard.edu/abs/2022MNRAS.514..535S} {514, 535}

\bibitem[\protect\citeauthoryear{{Sellek}, {Grassi}, {Picogna}, {Rab}, {Clarke}  \& {Ercolano}}{{Sellek} et~al.}{2024}]{Sellek24}
{Sellek} A.~D.,  {Grassi} T.,  {Picogna} G.,  {Rab} C.,  {Clarke} C.~J.,   {Ercolano} B.,  2024, \mn@doi [\aap] {10.1051/0004-6361/202450171}, \href {https://ui.adsabs.harvard.edu/abs/2024A&A...690A.296S} {690, A296}

\bibitem[\protect\citeauthoryear{{Shakura} \& {Sunyaev}}{{Shakura} \& {Sunyaev}}{1973}]{Shak}
{Shakura} N.~I.,  {Sunyaev} R.~A.,  1973, \aap, \href {http://adsabs.harvard.edu/abs/1973A%26A....24..337S} {24, 337}

\bibitem[\protect\citeauthoryear{{Sierra}, {Lizano}, {Mac{\'\i}as}, {Carrasco-Gonz{\'a}lez}, {Osorio}  \& {Flock}}{{Sierra} et~al.}{2019}]{Sierra19}
{Sierra} A.,  {Lizano} S.,  {Mac{\'\i}as} E.,  {Carrasco-Gonz{\'a}lez} C.,  {Osorio} M.,   {Flock} M.,  2019, \mn@doi [\apj] {10.3847/1538-4357/ab1265}, \href {https://ui.adsabs.harvard.edu/abs/2019ApJ...876....7S} {876, 7}

\bibitem[\protect\citeauthoryear{{Simon}, {Beckwith}  \& {Armitage}}{{Simon} et~al.}{2012}]{Simon12}
{Simon} J.~B.,  {Beckwith} K.,   {Armitage} P.~J.,  2012, \mn@doi [\mnras] {10.1111/j.1365-2966.2012.20835.x}, \href {http://adsabs.harvard.edu/abs/2012MNRAS.422.2685S} {422, 2685}

\bibitem[\protect\citeauthoryear{{Simon}, {Armitage}, {Li}  \& {Youdin}}{{Simon} et~al.}{2016}]{Simon16}
{Simon} J.~B.,  {Armitage} P.~J.,  {Li} R.,   {Youdin} A.~N.,  2016, \mn@doi [\apj] {10.3847/0004-637X/822/1/55}, \href {https://ui.adsabs.harvard.edu/abs/2016ApJ...822...55S} {822, 55}

\bibitem[\protect\citeauthoryear{{Standing} et~al.,}{{Standing} et~al.}{2023}]{Standing23}
{Standing} M.~R.,  et~al., 2023, \mn@doi [Nature Astronomy] {10.1038/s41550-023-01948-4}, \href {https://ui.adsabs.harvard.edu/abs/2023NatAs...7..702S} {7, 702}

\bibitem[\protect\citeauthoryear{{Steinacker} \& {Papaloizou}}{{Steinacker} \& {Papaloizou}}{2002}]{SteinackerPapaloizou2002}
{Steinacker} A.,  {Papaloizou} J.~C.~B.,  2002, \mn@doi [\apj] {10.1086/339892}, \href {http://adsabs.harvard.edu/abs/2002ApJ...571..413S} {571, 413}

\bibitem[\protect\citeauthoryear{{Su}, {Xie}, {Zhou}  \& {Thebault}}{{Su} et~al.}{2021}]{Su21}
{Su} X.-N.,  {Xie} J.-W.,  {Zhou} J.-L.,   {Thebault} P.,  2021, \mn@doi [\aj] {10.3847/1538-3881/ac2ba3}, \href {https://ui.adsabs.harvard.edu/abs/2021AJ....162..272S} {162, 272}

\bibitem[\protect\citeauthoryear{{Thebault} \& {Bonanni}}{{Thebault} \& {Bonanni}}{2025}]{Thebault25}
{Thebault} P.,  {Bonanni} D.,  2025, \mn@doi [\aap] {10.1051/0004-6361/202555457}, \href {https://ui.adsabs.harvard.edu/abs/2025A&A...700A.106T} {700, A106}

\bibitem[\protect\citeauthoryear{{Th{\'e}bault}, {Marzari}, {Scholl}, {Turrini}  \& {Barbieri}}{{Th{\'e}bault} et~al.}{2004}]{Thebault04}
{Th{\'e}bault} P.,  {Marzari} F.,  {Scholl} H.,  {Turrini} D.,   {Barbieri} M.,  2004, \mn@doi [\aap] {10.1051/0004-6361:20040514}, \href {https://ui.adsabs.harvard.edu/abs/2004A&A...427.1097T} {427, 1097}

\bibitem[\protect\citeauthoryear{{Th{\'e}bault}, {Marzari}  \& {Scholl}}{{Th{\'e}bault} et~al.}{2006}]{Thebault06}
{Th{\'e}bault} P.,  {Marzari} F.,   {Scholl} H.,  2006, \mn@doi [\icarus] {10.1016/j.icarus.2006.01.022}, \href {https://ui.adsabs.harvard.edu/abs/2006Icar..183..193T} {183, 193}

\bibitem[\protect\citeauthoryear{{Th{\'e}bault}, {Marzari}  \& {Scholl}}{{Th{\'e}bault} et~al.}{2008}]{Thebault08}
{Th{\'e}bault} P.,  {Marzari} F.,   {Scholl} H.,  2008, \mn@doi [\mnras] {10.1111/j.1365-2966.2008.13536.x}, \href {https://ui.adsabs.harvard.edu/abs/2008MNRAS.388.1528T} {388, 1528}

\bibitem[\protect\citeauthoryear{{Th{\'e}bault}, {Marzari}  \& {Scholl}}{{Th{\'e}bault} et~al.}{2009}]{Thebault09}
{Th{\'e}bault} P.,  {Marzari} F.,   {Scholl} H.,  2009, \mn@doi [\mnras] {10.1111/j.1745-3933.2008.00590.x}, \href {https://ui.adsabs.harvard.edu/abs/2009MNRAS.393L..21T} {393, L21}

\bibitem[\protect\citeauthoryear{{Trapman}, {Rosotti}, {Bosman}, {Hogerheijde}  \& {van Dishoeck}}{{Trapman} et~al.}{2020}]{Trapman20}
{Trapman} L.,  {Rosotti} G.,  {Bosman} A.~D.,  {Hogerheijde} M.~R.,   {van Dishoeck} E.~F.,  2020, \mn@doi [\aap] {10.1051/0004-6361/202037673}, \href {https://ui.adsabs.harvard.edu/abs/2020A&A...640A...5T} {640, A5}

\bibitem[\protect\citeauthoryear{{Vicente}, {Bern{\'e}}, {Tielens}, {Hu{\'e}lamo}, {Pantin}, {Kamp}  \& {Carmona}}{{Vicente} et~al.}{2013}]{Vicente13}
{Vicente} S.,  {Bern{\'e}} O.,  {Tielens} A.~G.~G.~M.,  {Hu{\'e}lamo} N.,  {Pantin} E.,  {Kamp} I.,   {Carmona} A.,  2013, \mn@doi [\apjl] {10.1088/2041-8205/765/2/L38}, \href {https://ui.adsabs.harvard.edu/abs/2013ApJ...765L..38V} {765, L38}

\bibitem[\protect\citeauthoryear{{Wang} \& {Goodman}}{{Wang} \& {Goodman}}{2017}]{Wang17}
{Wang} L.,  {Goodman} J.,  2017, \mn@doi [\apj] {10.3847/1538-4357/aa8726}, \href {https://ui.adsabs.harvard.edu/abs/2017ApJ...847...11W} {847, 11}

\bibitem[\protect\citeauthoryear{{Wang}, {Fischer}, {Horch}  \& {Xie}}{{Wang} et~al.}{2015}]{Wang15b}
{Wang} J.,  {Fischer} D.~A.,  {Horch} E.~P.,   {Xie} J.-W.,  2015, \mn@doi [\apj] {10.1088/0004-637X/806/2/248}, \href {https://ui.adsabs.harvard.edu/abs/2015ApJ...806..248W} {806, 248}

\bibitem[\protect\citeauthoryear{{Wilhelm}, {Portegies Zwart}, {Cournoyer-Cloutier}, {Lewis}, {Polak}, {Tran}  \& {Mac Low}}{{Wilhelm} et~al.}{2023}]{Wilhelm23}
{Wilhelm} M. J.~C.,  {Portegies Zwart} S.,  {Cournoyer-Cloutier} C.,  {Lewis} S.~C.,  {Polak} B.,  {Tran} A.,   {Mac Low} M.-M.,  2023, \mn@doi [\mnras] {10.1093/mnras/stad445}, \href {https://ui.adsabs.harvard.edu/abs/2023MNRAS.520.5331W} {520, 5331}

\bibitem[\protect\citeauthoryear{{Winn} \& {Fabrycky}}{{Winn} \& {Fabrycky}}{2015}]{Winn15}
{Winn} J.~N.,  {Fabrycky} D.~C.,  2015, \mn@doi [\araa] {10.1146/annurev-astro-082214-122246}, \href {https://ui.adsabs.harvard.edu/abs/2015ARA&A..53..409W} {53, 409}

\bibitem[\protect\citeauthoryear{{Winter}, {Haworth}, {Coleman}  \& {Nayakshin}}{{Winter} et~al.}{2022}]{Winter22}
{Winter} A.~J.,  {Haworth} T.~J.,  {Coleman} G. A.~L.,   {Nayakshin} S.,  2022, \mn@doi [\mnras] {10.1093/mnras/stac1564}, \href {https://ui.adsabs.harvard.edu/abs/2022MNRAS.515.4287W} {515, 4287}

\bibitem[\protect\citeauthoryear{{Yang}, {Johansen}  \& {Carrera}}{{Yang} et~al.}{2017}]{Yang17}
{Yang} C.~C.,  {Johansen} A.,   {Carrera} D.,  2017, \mn@doi [\aap] {10.1051/0004-6361/201630106}, \href {https://ui.adsabs.harvard.edu/abs/2017A&A...606A..80Y} {606, A80}

\bibitem[\protect\citeauthoryear{{Youdin} \& {Goodman}}{{Youdin} \& {Goodman}}{2005}]{Youdin05}
{Youdin} A.~N.,  {Goodman} J.,  2005, \mn@doi [\apj] {10.1086/426895}, \href {https://ui.adsabs.harvard.edu/abs/2005ApJ...620..459Y} {620, 459}

\bibitem[\protect\citeauthoryear{{Youdin} \& {Lithwick}}{{Youdin} \& {Lithwick}}{2007}]{Youdin07}
{Youdin} A.~N.,  {Lithwick} Y.,  2007, \mn@doi [\icarus] {10.1016/j.icarus.2007.07.012}, \href {https://ui.adsabs.harvard.edu/abs/2007Icar..192..588Y} {192, 588}

\bibitem[\protect\citeauthoryear{{Zagaria}, {Rosotti}  \& {Lodato}}{{Zagaria} et~al.}{2021}]{Zagaria21}
{Zagaria} F.,  {Rosotti} G.~P.,   {Lodato} G.,  2021, \mn@doi [\mnras] {10.1093/mnras/stab985}, \href {https://ui.adsabs.harvard.edu/abs/2021MNRAS.504.2235Z} {504, 2235}

\bibitem[\protect\citeauthoryear{{Zhu}, {Stone}, {Rafikov}  \& {Bai}}{{Zhu} et~al.}{2014}]{ZhuStoneBai2014}
{Zhu} Z.,  {Stone} J.~M.,  {Rafikov} R.~R.,   {Bai} X.-n.,  2014, \mn@doi [\apj] {10.1088/0004-637X/785/2/122}, \href {http://adsabs.harvard.edu/abs/2014ApJ...785..122Z} {785, 122}

\bibitem[\protect\citeauthoryear{{Ziegler}, {Tokovinin}, {Brice{\~n}o}, {Mang}, {Law}  \& {Mann}}{{Ziegler} et~al.}{2020}]{Ziegler20}
{Ziegler} C.,  {Tokovinin} A.,  {Brice{\~n}o} C.,  {Mang} J.,  {Law} N.,   {Mann} A.~W.,  2020, \mn@doi [\aj] {10.3847/1538-3881/ab55e9}, \href {https://ui.adsabs.harvard.edu/abs/2020AJ....159...19Z} {159, 19}

\bibitem[\protect\citeauthoryear{{van Terwisga} \& {Hacar}}{{van Terwisga} \& {Hacar}}{2023}]{VanTerwisga23}
{van Terwisga} S.~E.,  {Hacar} A.,  2023, \mn@doi [\aap] {10.1051/0004-6361/202346135}, \href {https://ui.adsabs.harvard.edu/abs/2023A&A...673L...2V} {673, L2}

\makeatother
\end{thebibliography}

\label{lastpage}
\end{document}